\documentclass[usletter,fleqn,usenatbib]{mnras}

\usepackage{newtxtext,newtxmath}

\usepackage[T1]{fontenc}
\usepackage{ae,aecompl}

\usepackage{graphicx}	
\usepackage{amsmath}	
\usepackage{amssymb}	

\newcommand{\be}{\begin{equation}}
\newcommand{\ee}{\end{equation}}

\title[Gas Velocity Dispersions and SFRs in FIRE Disks]{Swirls of FIRE: Spatially Resolved Gas Velocity Dispersions and Star Formation Rates in FIRE-2 Disk Environments}

\author[M. E. Orr et al.]{Matthew E. Orr$^{1}$\thanks{E-mail: meorr@caltech.edu},  
Christopher C. Hayward$^{2}$, 
Anne M. Medling$^{3,4}$\thanks{Hubble Fellow}, \newauthor
Philip F. Hopkins$^{1}$,
Norman Murray$^5$,
Jorge L. Pineda$^6$, \newauthor
 Claude-Andr\'e Faucher-Gigu\`{e}re$^{7}$, 
 Du\v{s}an Kere\v{s}$^{8}$,  
and Kung-Yi Su$^{1,2}$\\
$^{1}$TAPIR, Mailcode 350-17, California Institute of Technology, Pasadena, CA 91125, USA\\
$^{2}$Center for Computational Astrophysics, Flatiron Institute, 162 Fifth Avenue, New York, NY 10010, USA\\
$^{3}$Ritter Astrophysical Research Center University of Toledo Toledo, OH 43606, USA\\
$^{4}$Research School for Astronomy \& Astrophysics Australian National University Canberra, ACT 2611, Australia\\
$^{5}$Canadian Institute for Theoretical Astrophysics, 60 St George Street, University of Toronto, ON M5S 3H8, Canada\\
$^{6}$Jet Propulsion Laboratory, California Institute of Technology, 4800 Oak Grove Drive, Pasadena, CA 91109-8099, USA\\
$^{7}$Department of Physics and Astronomy and CIERA, Northwestern University, 2145 Sheridan Road, Evanston, IL 60208, USA\\
$^{8}$Department of Physics, Center for Astrophysics and Space Science, University of California at San Diego, 9500 Gilman Drive, \\ La Jolla, CA 92093, USA\\
}
\date{Accepted XXX. Received YYY; in original form ZZZ}

\pubyear{2019}

\begin{document}
\label{firstpage}
\pagerange{\pageref{firstpage}--\pageref{lastpage}}
\maketitle

\begin{abstract}
We study the spatially resolved (sub-kpc) gas velocity dispersion ($\sigma$)--star formation rate (SFR) relation in the FIRE-2 (Feedback in Realistic Environments) cosmological simulations.  We specifically focus on Milky Way mass disk galaxies at late times.  In agreement with observations, we find a relatively flat relationship, with $\sigma \approx 15-30$ km/s in neutral gas across 3 dex in SFRs.  We show that higher dense gas fractions (ratios of dense gas to neutral gas) and SFRs are correlated at constant $\sigma$. Similarly, lower gas fractions (ratios of gas to stellar mass) are correlated with higher $\sigma$ at constant SFR. The limits of the $\sigma$-$\Sigma_{\rm SFR}$ relation correspond to the onset of strong outflows.   We see evidence of ``on-off'' cycles of star formation in the simulations, corresponding to feedback injection timescales of 10-100 Myr, where SFRs oscillate about equilibrium SFR predictions. Finally, SFRs and velocity dispersions in the simulations agree well with feedback-regulated and marginally stable gas disk (Toomre's $Q =1$) model predictions, and the data effectively rule out models assuming that gas turns into stars at (low) constant efficiency (i.e., 1\% per free-fall time).  And although the simulation data do not entirely exclude gas accretion/gravitationally powered turbulence as a driver of $\sigma$, it appears to be strongly subdominant to stellar feedback in the simulated galaxy disks.
\end{abstract}




\section{Introduction}\label{sec:intro}

Star formation in the local universe (at $z \approx 0$) is dominated by spiral galaxies of approximately Milky Way mass \citep[stellar masses of $\sim 10^{10} - 10^{11}$ M$_\odot$,][]{Brinchmann2004, Behroozi2013}.  A hallmark of these galaxies at late times is relatively constant steady star formation in the galaxy disk over the last several billion years \citep{Ma2017, Simons2017}. Understanding how star formation and the interstellar medium (ISM) interact in disk environments is therefore crucial to understanding how these galaxies have evolved, and continue to grow. 

Observationally, star formation in galaxies has often been studied through the lens of the Kennicutt-Schmidt (KS) and Elmegreen-Silk (ES) relations, both empirical star formation scaling relations relating gas surface densities $\Sigma_{\rm gas}$, and gas surface densities times the local orbital dynamical time $\Sigma_{\rm gas}\Omega$, to the star formation surface density $\Sigma_{\rm SFR}$, respectively, to probe the connections between local gas masses and star formation (\citealt{Kennicutt1989}, and see \citealt{Kennicutt2012} for a recent review).  The scatter seen in the KS and ES relations have long been viewed as ``the weather'' of variations in local conditions and differing resulting star formation equilibria \citep{Bigiel2008, Leroy2008, Leroy2013}.  However, the KS and ES relations themselves do not directly speak to the \emph{dynamical} state of the gas in galaxies, relating only the surface densities of gas to star formation rates.  And so theoretical work has been afforded significant leeway in explaining the origins of these star formation scaling relations, from being the direct consequence of stellar feedback, to being purely a result of gas dynamics and migration in galaxies \citep{Faucher-Giguere2013, Krumholz2016}.

Theoretical work has shown these relations can be understood on average by treating star formation as a self-regulating process \citep{Thompson2005, Ostriker2011, Faucher-Giguere2013, Hayward2017}.  The hierarchy of timescales involved in galaxies provided motivation for this: star formation, and the bulk of stellar feedback, occurs on the timescale of a few million years or less (tens at most), but the dynamical times of galaxies can be on the order of 100 Myr.  If galaxies are to have long-lived coherent structures (e.g., spiral arms), star formation must either be on-average in equilibrium with those structures, or be unable to greatly affect them. 


Recent work exploring star formation as a non-equilibrium process on the scales of giant molecular clouds (GMCs, $\gtrsim100$ pc) has argued for the importance of local gas dynamics (e.g., gas turbulence) in producing both the scatter and normalization in star formation scaling relations like Kennicutt-Schmidt \citep{Benincasa2016, Torrey2016, Sparre2017, Orr2019}. Work by \citet{Torrey2016} to understand how the outflows and ``breathing modes'' of central molecular regions are tied to star formation shows that non-equilibrium star formation rates, on timescales of tens of millions of years, arise naturally from the competition between the stellar feedback and dissipative dynamical processes.   Further works by \citet{Benincasa2016} and \citet{Orr2019} have explored the dynamical and star formation rate responses of the ISM in disk environments to changing velocity dispersions and feedback injection, on the timescales of tens of Myr.

Much work on star formation in both equilibrium and non-equilibrium contexts has focused on the stability of gas in galaxies against gravitational fragmentation and collapse.  Gas velocity dispersions $\sigma$ are seen as a key driver of gravitational stability and a measure of the dynamical state of gas in galaxy disks, with dense gas velocity dispersions (observationally, line-widths from molecular gas tracers like CO or HCN) being tied to the degree of turbulent support within the disks.  The connection between $\sigma$ and stability is most explicit when gas stability is calculated using the (modified) Toomre-Q parameter \citep{Toomre1964}, having a form similar to:
\be \label{eq:Q}
\tilde Q_{\rm gas} \equiv  \frac{\sqrt{2}\sigma_z \Omega}{\pi G (\Sigma_{\rm gas} + \gamma\Sigma_\star)} \; , 
\ee
where $\sigma_z$ is the vertical (line-of-sight) velocity dispersion in the gas\footnote{It is often assumed that the turbulence field is roughly isotropic on a disk scale height, so the in-plane turbulence in a disk (used for the `classical' Toomre-Q parameter) is equivalent to the line-of-sight component, thus $\sigma_R \approx \sigma_z$.}, $\Omega$ the dynamical angular velocity ($\Omega \equiv v_c/R$), and $ (\Sigma_{\rm gas} + \gamma\Sigma_\star)$ being the effective disk surface density, where $\Sigma_{\rm gas}$ is the neutral (atomic + molecular) gas surface density and the $\gamma \Sigma_\star$ represents the fraction of the stellar component within the gas disk scale height (and thus contributing to the self-gravity of the disk)\footnote{For all subsequent calculations of $\tilde Q_{\rm gas}$ in this work, we calculate the $\gamma$ factor assuming that the stellar component has an exponential scale height, and thus is $\gamma = 1 -\exp{(-\sigma_{z,{\rm gas}}/\sigma_{z,\star})}$, to simplify direct calculations of this factor in comparisons with observations.}.    

For gas in galaxies with $\tilde Q_{\rm gas} \lesssim 1$, there is insufficient turbulent support ($\sigma$) to prevent fragmentation and gravitational collapse.  This gas collapses to form stars, removing it from the gas reservoir.  In the feedback-regulated framework, those stars then inject feedback to the remaining gas to stabilize it.  On the other hand, gas with $\tilde Q_{\rm gas} \gg 1$ is likely to be dynamically expanding or in the midst of an outflow event.  In the event that this gas is not being actively driven to larger values of $\tilde Q_{\rm gas}$, it is expected that the gas dissipates its turbulence rapidly on a disk crossing time, driving to $\tilde Q_{\rm gas} = 1$.  And so, for the purposes of supersonically turbulent disks, we expect that gas ought to converge to a $\tilde Q_{\rm gas} \approx 1$.

The rate at which marginally unstable gas ($\tilde Q_{\rm gas} \lesssim 1$) in galaxies ought to form stars is theoretically uncertain.  Some feedback-regulation arguments, for example that feedback from young stars balances the gravitational weight of the ISM, are completely agnostic to the velocity dispersions or specific value of Toomre-Q \citep{Ostriker2011}.  And so, in spite of good arguments for $\tilde Q_{\rm gas} \rightarrow 1$ convergence on large scales, we are still left with conflicting evidence and predictions for how gas turbulence (velocity dispersions/observed line-widths) is sourced and whether it is a direct consequence, or cause of, local star formation rates (see \S~\ref{subsec:scalings} for a more in-depth discussion).  That of the quantities involved in predicting $\tilde Q_{\rm gas}$, only $\sigma$ appreciably evolves on timescales $\lesssim 100$ Myr (barring significant gas inflows/outflows) highlights the importance of understanding how velocity dispersions and star formation connect.

Much of the work understanding the velocity dispersion structure of gas in local galaxies and their relationship with local star formation rates has investigated the H{\scriptsize II} regions of those galaxies \citep{Larson1981, Gallagher1983, Rozas1998, Rozas2006, Zhou2017}, due in part to the relative difficulty in adequately measuring at high velocity resolution the velocity structure of the colder, fainter dense molecular gas tracers like CO or HCN. Modern surveys are just now beginning to report on the galaxy-wide, spatially resolved velocity structure and surface density distributions in the cold, dense molecular gas \citep{Leroy2017, Gallagher2018,Gallagher2018a, Sun2018, Querejeta2019}.  Recent work has focused on how dense gas fractions and the mid-plane pressure of the ISM correlate with star formation efficiencies, and whether or not variations in these quantities can explain the variations seen in star formation rates across the galaxies as a whole.

Cosmological zoom-in simulations are now beginning to resolve the ISM on scales within GMC complexes, with sub-parsec spatial resolution and mass resolutions reaching sub-$10^3$ M$_\odot$ \citep{Hopkins2014, Wetzel2016, Hopkins2018:fire}.  Given the ability to resolve ISM dynamics on the scales within star forming regions, recent work by \citet{Orr2018} has explored the spatially resolved properties of the Kennicutt-Schmidt relation and its dependences on local gas properties in the FIRE simulations.  Another investigation of the FIRE suite by \citet{El-Badry2018} investigated the kinematic and morphological properties of gas within galaxies.  However, neither study explicitly explored the connection between the spatially resolved gas kinematics (velocity dispersions) and the local star formation rates.

In this work, we explore the relationship between various spatially resolved measures of gas velocity dispersion and star formation rates using the FIRE-2 cosmological zoom-in simulations.  We explore the dependences of the velocity dispersions and star formation rates in Milky Way-like disk environments on various local gas properties, disk conditions, and recent star formation histories.  We compare the FIRE simulations where possible with spatially resolved observational datasets.  Through these means, we test the predictions of several classes of star formation theories that connect gas velocity dispersions/supersonic turbulence with local star formation rates and gas stability.

\section{Simulations \& Analysis Methods}

\subsection{FIRE-2 Simulations}\label{sec:sims}

\begin{table}\caption{Summary of $z\approx0$ properties of the FIRE-2 Milky Way-like galaxies used in this work}\label{table:galprops}
\begin{tabular}{lccccc}
\hline
Name & $\log(\frac{M_\star}{{\rm M_\odot}})$ & $\log(\frac{M_{\rm gas}}{{\rm M_\odot}})$ & $\frac{R_{\star,1/2}}{\rm kpc}$& $\frac{R_{\rm gas,1/2}}{\rm kpc}$ & $\frac{v_c}{\rm km/s}$* \\
 \hline
m12b  & 10.8 & 10.3 & 2.7 & 9.4 & 266 \\
m12c  & 10.7 & 10.3 & 3.4 & 8.6 & 232 \\
m12f   & 10.8 & 10.4 & 4.0 & 11.6 & 248 \\
m12i   & 10.7 & 10.3 & 2.9 & 9.8 & 232 \\
m12m & 10.9 & 10.4 & 5.6 & 10.2 & 283 \\
m12r   & 10.2 & 10.0 & 4.7 & 9.9 & 156 \\
m12w  & 10.6 & 9.8 & 3.1 & 3.1& 244 \\
 \hline
\multicolumn{6}{l}{Note: all quantities measured within a 30~kpc cubic aperture.}\\
\multicolumn{6}{l}{*Circular velocities evaluated at $R_{\rm gas,1/2}$.}\\
\end{tabular}
\end{table}
The simulations used here are the FIRE-2 Milky Way-mass galaxies presented in \citep{Hopkins2018:fire}.  They were run with the gravity+hydrodynamics code {\scriptsize GIZMO} \citep{Hopkins2015:gizmo}.  Specifically, the simulations were run with a mesh-free Lagrangian Godunov (meshless finite mass, MFM) method.  A brief summary of the $z \approx 0$ properties of the galaxy simulations are included in Table~\ref{table:galprops}.  These simulations all have a minimum baryonic particle mass of $m_{b,min} = 7100$ M$_\odot$, and minimum adaptive force softening lengths $<$1~pc.  As the softening lengths are adaptive, it is useful to note that the typical/median softening length within the disk in one of the runs at $z=0$, \textbf{m12i}, is $h\sim 20-40$ pc (at $n\sim 1$ cm$^{-3}$). The minimum length scales considered in this work are hundreds of parsecs, and so are about three orders of magnitude above the minimum resolvable scales in the simulations. All of the simulations employ a standard flat $\Lambda$CDM cosmology with $h \approx 0.7$, $\Omega_M = 1 - \Omega_\Lambda \approx 0.27$, and $\Omega_b \approx 0.046$.

The star formation prescription in the FIRE-2 simulations is described in more detail in \citet{Hopkins2018:fire}, however, to reiterate here: star formation occurs on a free-fall time in gas which is dense ($n >10^3$ cm$^{-3}$), molecular (per the \citet{Krumholz2011} empirical fit for molecular gas fractions as a function of local gas column density), self-gravitating (viral parameter $\alpha_{\rm vir} < 1$) and Jeans-unstable below the resolution scale.
Once a star particle is formed, it is treated as a single stellar population with known age, metallicity, and mass.   We explicitly follow feedback from: supernovae (Type Ia and II), stellar mass loss (OB/AGB-star winds), photoionization and photoelectric heating, and radiation pressure.  Detailed descriptions of these physics and their implementation can be found in \citet{Hopkins2018:fire}.  All feedback quantities are taken from standard stellar population models ({\scriptsize STARBURST99}, \citealt{Leitherer1999}), assuming a \citealt{Kroupa2002} IMF.

\subsection{Mapping out FIRE-2: Resolved `Observations' of the Simulations}\label{sec:maps}

We generate mock observational maps from a set of snapshots with $z\lesssim 0.1$ in the manner of \citet{Orr2018}, by  projecting the galaxies face-on with respect to the stellar angular momentum, then binning star particles and gas elements into square pixels\footnote{In our analysis we treat pixels from all simulations and all times equally, unless otherwise stated.} with side-lengths (\emph{i.e.}, ``pixel sizes'') 250 -- 750~pc.  The maps themselves are 30~kpc on a side, integrating gas and stars within $\pm 15$ kpc of the galaxy midplane (to exclude line-of-sight projections of other galaxies within the larger cosmological box).  

We generate star formation rate tracers analogous to observational measures of star formation by calculating the average star formation rates over the past 10 and 100 Myr using the star particle ages, with,
\be
\Sigma_{\rm SFR}^{\Delta t} = \frac{M_\star({\rm age} < \Delta t)}{\eta l^2 \Delta t} \; ,
\ee
where $M_\star({\rm age} < \Delta t)$ is the summed mass of all star particles in the pixel with ages less than the averaging window $\Delta t$, $l^2$ is the pixel size (in kpc$^2$), and $\eta$ is a factor correcting for mass loss from stellar winds and evolutionary effects using predictions from {\scriptsize STARBURST99} \citep{Leitherer1999}, with values of 0.85 and 0.70 for the $\Delta t =$10 Myr and 100 Myr timescales, respectively.  The 10 and 100 Myr intervals were chosen for their approximate correspondence with the timescales traced by recombinations lines like H$\alpha$, and emission in the UV or FIR \citep{Kennicutt2012}\footnote{Post-processing the snapshots to explicitly model H$\alpha$ or UV fluxes would make for a more direct comparison to observations, but accounting for, \emph{e.g.}, dust and other complexities is beyond the scope of this work, where we wish to focus on the ``true'' SFRs.}.  The instantaneous star formation rate of the gas is also considered. 

We calculate column densities and mass-averaged line-of-sight velocity dispersions $\sigma_z$ for (1) the total neutral gas column $\Sigma_{\rm gas}$, and (2) the ``cold and dense'' gas $\Sigma_{\rm C\&D}$  with $T < 500$~K and $n_{\rm H} > 1$ cm$^{-3}$.  The latter gas reservoir taken as a proxy for the cold molecular gas in the simulations following the methodology of \citet{Orr2018} but with a more liberal higher temperature cut for what constitutes ``cold'' in the ISM.  These roughly correspond to the observed velocity dispersions in the H{\scriptsize I} $+$ H$_2$ or H$_2$ gas, respectively.  We calculate the local gas velocity dispersion in each pixel as the mass-weighted standard deviation of the line-of-sight velocities of the qualifying gas elements (for each type of velocity dispersion).  This is identical to taking the width of a (maximum likelihood) fitted single-component Gaussian.



We also estimate the angular velocity $\Omega$ in each pixel, defined here as
\begin{equation}\label{eqn:omega}
\begin{centering}
\Omega = \frac{v_c}{R} = \frac{(GM(<R))^{1/2}}{R^{3/2}} \; ,
\end{centering}
\end{equation}
where R is the galactocentric radius of the pixel, and M(< R) is the total mass enclosed within a sphere of that radius. 




We compare our simulations with various observational studies of spatially resolved gas line-of-sight velocity dispersions in star-forming galaxies. \citet{Zhou2017} provide a dataset at $\sim$kpc scales from the SAMI Galaxy Survey \citep{Croom2012, Scott2018}, relating H{\scriptsize II} region velocity dispersions to H$\alpha$-inferred SFRs.  We compare these data to our neutral gas velocity dispersion and 10 Myr-averaged SFR tracers in 750~pc pixels.  With our higher-resolution pixel at 250 pc, we take data from \citet[][H{\scriptsize II} region velocity dispersions]{Rozas2006} for comparison with neutral gas velocity dispersions in FIRE, and from \citet[][HCN-traced gas velocity dispersions and dense gas depletion times]{Querejeta2019} for comparison with our cold and dense gas.

\section{Spatially Resolved Velocity Structure and SFRs: Comparing Simulations \& Observations}\label{sec:results}
\begin{figure*}
	\centering
	\includegraphics[width=0.97\textwidth]{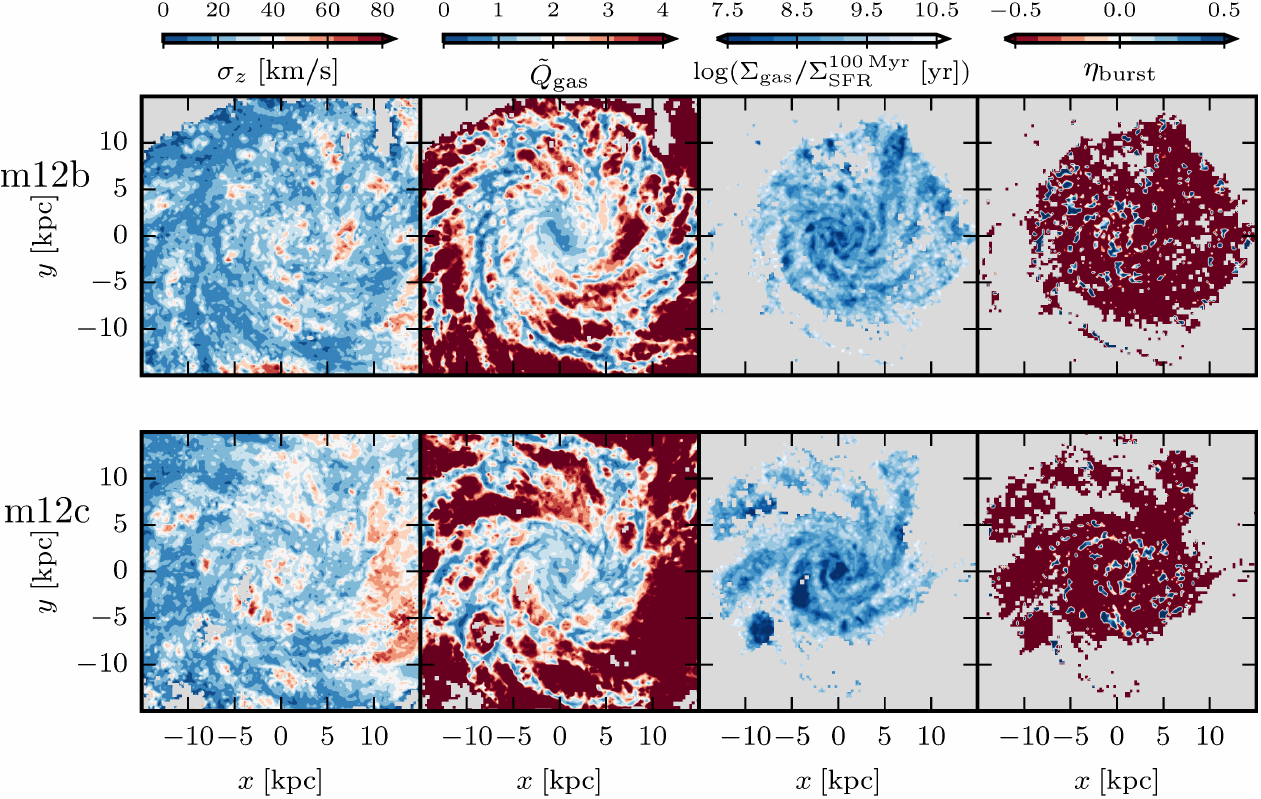}
	\caption{Face-on maps of star formation and gas quantities in two of the FIRE galaxies (\textbf{rows}: \textbf{m12b} and \textbf{m12c}, see Fig.~\ref{fig:galsm12m-w} for other galaxy simulations) at $z=0$ with 250 pc pixel size. Regions with fewer than $\sim10$ gas elements (or no star formation, in the cases of the right two columns) are excluded, and colored light grey. \textbf{Left:} Velocity dispersions (including inflow/outflow) in neutral (atomic + molecular) gas.  Spiral structures correspond to low $\sigma_z \lesssim 40$ km/s dispersions, with interspersed high-dispersion structures. \textbf{Center left:} Turbulent Toomre-Q parameter for gas, $\tilde Q_{\rm gas}$.  Bluer regions ($\tilde Q_{\rm gas} < 2$) are at least marginally unstable.  \textbf{Center right:} Gas depletion time averaged over 100 Myr. Several large ($\sim$kpc) regions with short depletion times correspond with large bubbles of high-$\tilde Q_{\rm gas}$, indicative of SNe super-bubbles. \textbf{Right:} ``Burstiness'' measure $\eta_{\rm burst} \equiv (\Sigma_{\rm SFR}^{\rm 10\,Myr}-\Sigma_{\rm SFR}^{\rm 100\,Myr})/(\Sigma_{\rm SFR}^{\rm 10\,Myr}+\Sigma_{\rm SFR}^{\rm 100\,Myr})$.  Redder (bluer) regions indicate rapidly decaying (rising) SFRs.  Much of the area of the galaxies are covered by regions with decaying local star formation, while small regions with locally rising SFRs correspond to GMCs.}
	\label{fig:galsm12b-i}
\end{figure*}

Figure~\ref{fig:galsm12b-i} shows the $z=0$ snapshots of two of the seven \textbf{m12} galaxies (by row), for a number of resolved physical quantities (by column) at a pixel size of 250 pc (the other five are found in Figure~\ref{fig:galsm12m-w} of Appendix~\ref{sec:appendix}).  Broadly speaking, the typical values of the velocity dispersions and star formation rates seen in the FIRE Milky Way-mass spirals agree well with similar resolved observations of local disk galaxies \citep{Rozas2006, Zhou2017, Querejeta2019}.  Across all seven of our \textbf{m12} galaxies, 95\% of the pixels (that also have had recent star formation according to either the 10 Myr or 100 Myr average star formation rate tracer) have velocity dispersions in the neutral gas of 10-50 km/s.

The galaxies are dominated by regions with low ($<40$ km/s, blue shades, first column) velocity dispersions, with pockets of high dispersion gas.  Although correlated by definition, it is not easy to see structures traced by the low or high values of Toomre-Q in velocity dispersions.  However, it \emph{is} easily seen in several cases that regions with the shortest depletion times (as calculated with the 100 Myr SFR tracer, third column) do correspond to regions of high Toomre-Q.  Lastly (fourth column), a measure of the ``burstiness'' of star formation $\eta_{\rm burst} = (\Sigma_{\rm SFR}^{\rm 10\,Myr}- \Sigma_{\rm SFR}^{\rm 100\, Myr})/(\Sigma_{\rm SFR}^{\rm 10 \, Myr}+ \Sigma_{\rm SFR}^{\rm 100\,Myr})$, shows that the galaxies are dominated in area by regions that have formed stars in the past 100 Myr (and thus, for the most part, those stars have injected all their feedback already) but have had little or no star formation in the past 10 Myr.  Given the small size of the regions with significant 10 Myr star formation rates, this result is somewhat smoothed on 750 pc scales. The fact that the 10 and 100 Myr average star formation rates are rarely equal, when considering 250 pc scales in galaxies, points to the fact that even in galaxies with consistent \emph{global} SFRs, star formation remains quite bursty locally.
\subsection{Various Tracers of Velocity Dispersion and Star Formation Rate Timescale}\label{sec:sfrs}

\begin{figure*}
	\centering
	\includegraphics[width=0.99\textwidth]{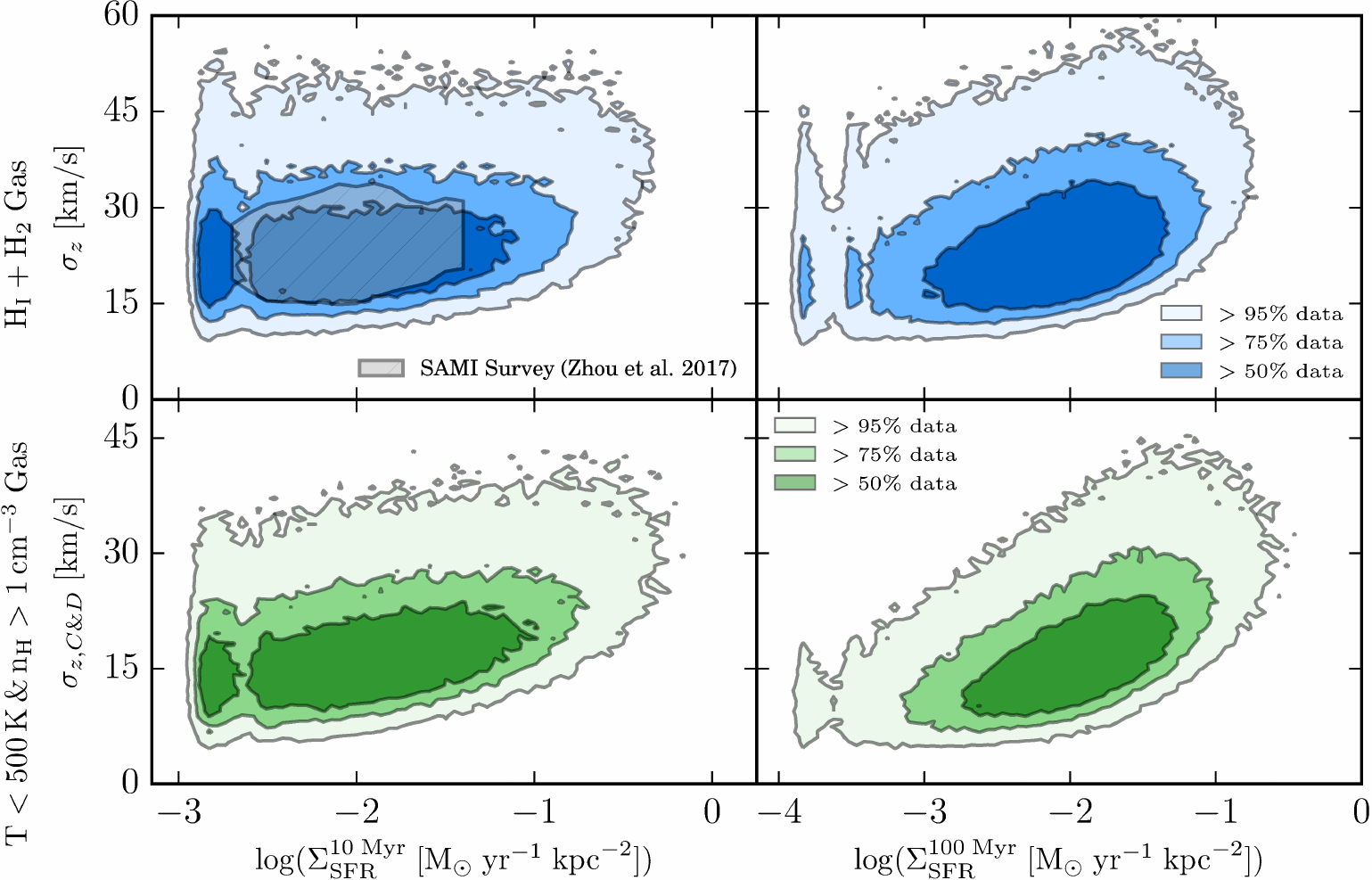}
	\caption{Distributions of spatially resolved (750 pc pixel size) line-of-sight gas velocity dispersions ($\sigma_z$) and SFR surface densities for various gas and SFR tracers in the Milky Way-mass FIRE simulations for $z \lesssim 0.1$.  Gas velocity dispersions are the mass-weighted standard deviation of the line-of-sight velocities in gas, intentionally including inflow, outflow, non-circular galactic motions, etc. Data are stacked together from all individual m12 galaxy simulations (see, Fig.~\ref{fig:galrelations}).  Filled contours indicate 95, 70, 50-percentile inclusion regions for the simulation data.  Velocity dispersions in neutral gas as a function of 10 Myr-averaged SFR are compared with observational data from \citet{Zhou2017}.  Across $\sim 3$ dex in SFRs, gas velocity dispersions are nearly constant, with a rising lower envelope of dispersions at a given SFR.  The velocity dispersions for the cold and dense gas ($T < 500$ K and $n > 1$ cm$^{-3}$, \emph{bottom row}) are lower than for neutral gas (atomic + molecular hydrogen, \emph{top row}), indicating the dynamically colder state of the dense molecular component of the ISM.  Tracers with longer averaging timescales (100 Myr vs. 10 Myr, \emph{right and left columns, respectively}) are able to trace the relation to lower star formation rates and gas velocity dispersions, showing that there is a trend, but that it is very weak and only apparent over longer averaging timescales.}
	\label{fig:tracers}
\end{figure*}

Figure~\ref{fig:tracers} shows the stacked data from all the snapshots of the Milky Way massed FIRE spirals. We can see the velocity dispersion--SFR relation in the simulations for variously weighted tracers of gas velocity dispersion and star formation rate (see Figure~\ref{fig:galrelations} in Appendix~\ref{sec:appendix} for individual galaxy $\sigma_z$--SFR distributions).  The extent of the data to low star formation rates in each panel is mass resolution-limited, with lower limits of $\Sigma_{\rm SFR} \approx 10^{-2.8},$ $10^{-3.75}$ M$_\odot$ yr$^{-1}$ kpc$^{-2}$ for the (750 pc)$^2$ pixels, with minimum baryonic masses of 7100 M$_\odot$ at 10 Myr and 100 Myr (with their associated evolutionally mass correction factors), respectively. 

Generally, the core of the distributions for the neutral (atomic + molecular) gas and the cold and dense gas velocity dispersions are between 15-40 km/s and 10-30 km/s, respectively.   There is a tail in the distributions to $\approx 60$ km/s and $\approx 45$ km/s for the 95\% data inclusion regions for their respective ISM components.  For the Milky Way-like rotational velocities $v_c \approx 240$ km/s of these simulations, the dispersion ratios are $\sigma_z/v_c \approx 0.06-0.25$ and $0.04-0.19$ in the gas (respective to $\sim 95\%$ of the two ISM components).  That is to say, the disks are thin in the FIRE \textbf{m12} simulations.

We compare directly with the SAMI IFU data of \citet{Zhou2017} in the top left panel of Figure~\ref{fig:tracers}, for our 10 Myr-averaged star formation rate and neutral gas velocity dispersion data.  Their data has complete overlap with our 75\% inclusion region (and nearly total overlap with our 50\% region).  Our inclusion of \textbf{m12w}, with its fairly `hot' disk is a large part of the spray to higher velocity dispersions in the FIRE data.  Additionally, as we orient our galaxies face-on, and thus have negligible beam-smearing or inclination effects in the line-of-sight velocity dispersions, we do not need to throw out pixels with velocity gradients as done in \citet{Zhou2017}.  Any velocity gradients within the face-on pixels thus correspond to disk structural properties, and warrant inclusion in our analysis.

The distributions all have increasing velocity dispersions as a function of star formation rates.  However, the effect is fairly weak, on the order of only a few km/s with the 10 My star formation rate tracer.   The most visible case, with the 100 Myr star formation rate tracer, amounts to $\sim 7$ km/s per dex in $\Sigma_{\rm SFR}$, effectively doubling the velocity dispersions across our dynamic range of star formation rates.  Universal, however, is the rising lower envelope of velocity dispersions as a function of star formation rate.  The nature of this lower envelope is explored in \S~\ref{subsec:outflows}.

\subsection{Correlations with Various Mass Surface Densities}
\begin{figure*}
	\centering
	\includegraphics[width=0.99\textwidth]{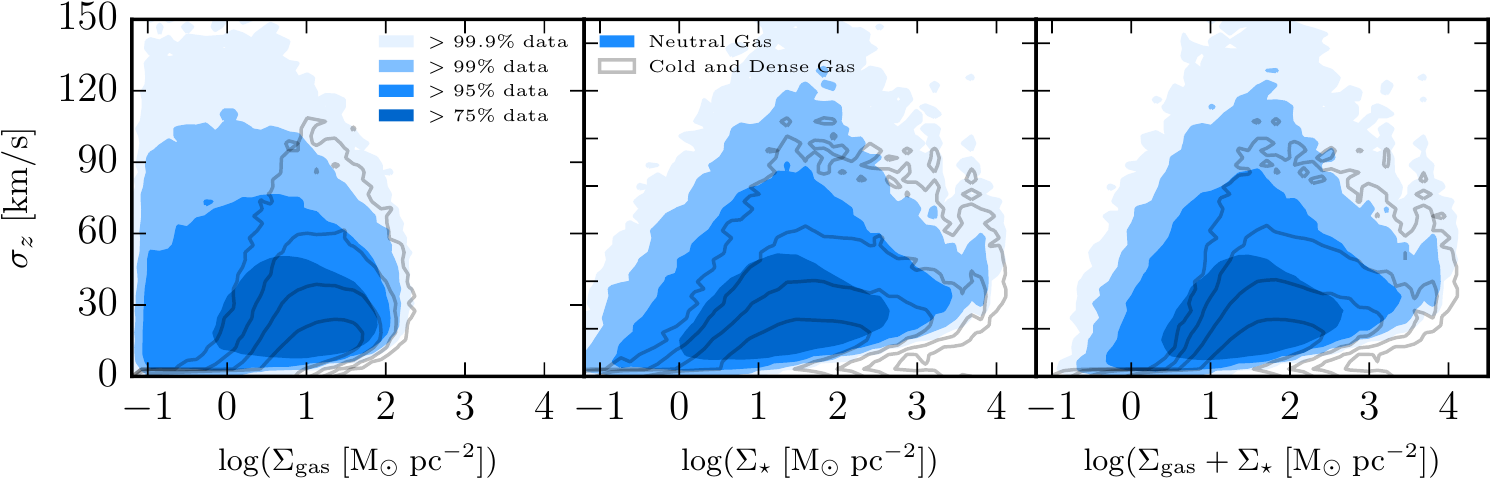}
	\caption{Distribution of $\sigma_z$ (as Fig.~\ref{fig:tracers}) versus gas surface density ($\Sigma_{\rm gas}$), stellar surface density ($\Sigma_\star$), and total surface density ($\Sigma_{\rm gas}+\Sigma_\star$).  Unfilled contours indicate velocity dispersions and gas surface densities of the cold and dense ($T <500$ K and $n > 1$ cm$^{-3}$) component, with identical data inclusion percentages.  We do not plot below below $\Sigma_{\rm gas} = 0.1$ M$_\odot$ pc$^{-2}$ to ensure at least $\sim 10$ gas elements per pixel for calculating $\sigma_z$.  \textbf{Left:} As neutral gas surface densities exceed $\sim$few M$_\odot$ pc$^{-2}$, the ISM transitions to a predominantly molecular form, and velocity dispersions rise in the cold ISM component. \textbf{Center:} Largest scatter in velocity dispersions occurs for $\log\Sigma_\star \sim 1.5$.  \textbf{Right:} Total surface density--velocity dispersion distribution is very similar to the stellar surface density--velocity distribution in neutral gas, but for cold and dense gas there is a steeper rise in dispersions (and generally cold gas content) around total surface densities of $\sim$10 M$_\odot$ pc$^{-2}$. Generally higher neutral gas surface densities have a lower scatter to high dispersions, whereas the cold and dense gas velocity dispersions consistently rise with increasing gas and stellar surface densities.}
	\label{fig:sigmasurfs}
\end{figure*}
Figure~\ref{fig:sigmasurfs} shows how the line-of-sight velocity dispersions relate to the various (gas, stellar, gas + stellar) mass surface densities in the galaxies.  The neutral (atomic + molecular) velocity dispersions are plotted in (blue) shaded contours, and the ``cold and dense'' gas velocity dispersions are plotted in unfilled, grey contours.  Generally, the neutral gas velocity dispersions exhibit less scatter at higher gas surface densities.  This lessened scatter may be explained by the fact that high surface density gas disks are more self-bound gravitationally, resulting in shorter gas scale heights and thus shorter eddy turnover times for a given $\sigma_z$.  If supersonic turbulence typically runs down on a eddy turnover time, then high velocity dispersions would be quickly dissipated in the ISM. For surface densities including the stellar component (center and right panels), the velocity dispersions peak in the range of $\Sigma_\star = 10 - 10^2$ M$_\odot$ pc$^{-2}$.  This peak may be an inflection point between velocity dispersions driving large scale heights and the self-gravity of the disks increasing the rate of turbulence dissipation in the ISM.

Like the velocity dispersion--SFR relations in Figure~\ref{fig:tracers}, the velocity dispersions have a lower envelope of dispersions for a given surface density.  This lower limit can be rationalized as a stability limit.  For dispersions below a certain value, the ISM would be gravitationally unstable to fragmentation and collapse.  In this case, that unstable ISM turns into stars, both removing that gas from the dispersion relation and also causing feedback that will drive turbulence in the remaining at-least-marginally stable gas.

For all mass surface densities, the ``cold and dense'' gas velocity dispersions generally rise with surface density, though always have a lower overall normalization compared to the neutral gas velocity dispersions.  The rising cold and dense gas dispersions are in line with the fact that the mass fraction of that gas phase is rising, and that it is a dynamically colder component of the ISM.  The rapid increase in velocity dispersions in the cold and dense gas between $\Sigma_{\rm gas} \approx$ 1-10 M$_\odot$ pc$^{-2}$ overlaps with the H{\scriptsize I} to H$_2$ transition threshold generally discussed in the literature \citep{Krumholz2008, Sternberg2014, Pineda2017}.

%
\subsection{Gas Fractions, Dense Gas Fractions, and Gas Stability}\label{sec:fractions}
\begin{figure*}
	\centering
	\includegraphics[width=\textwidth]{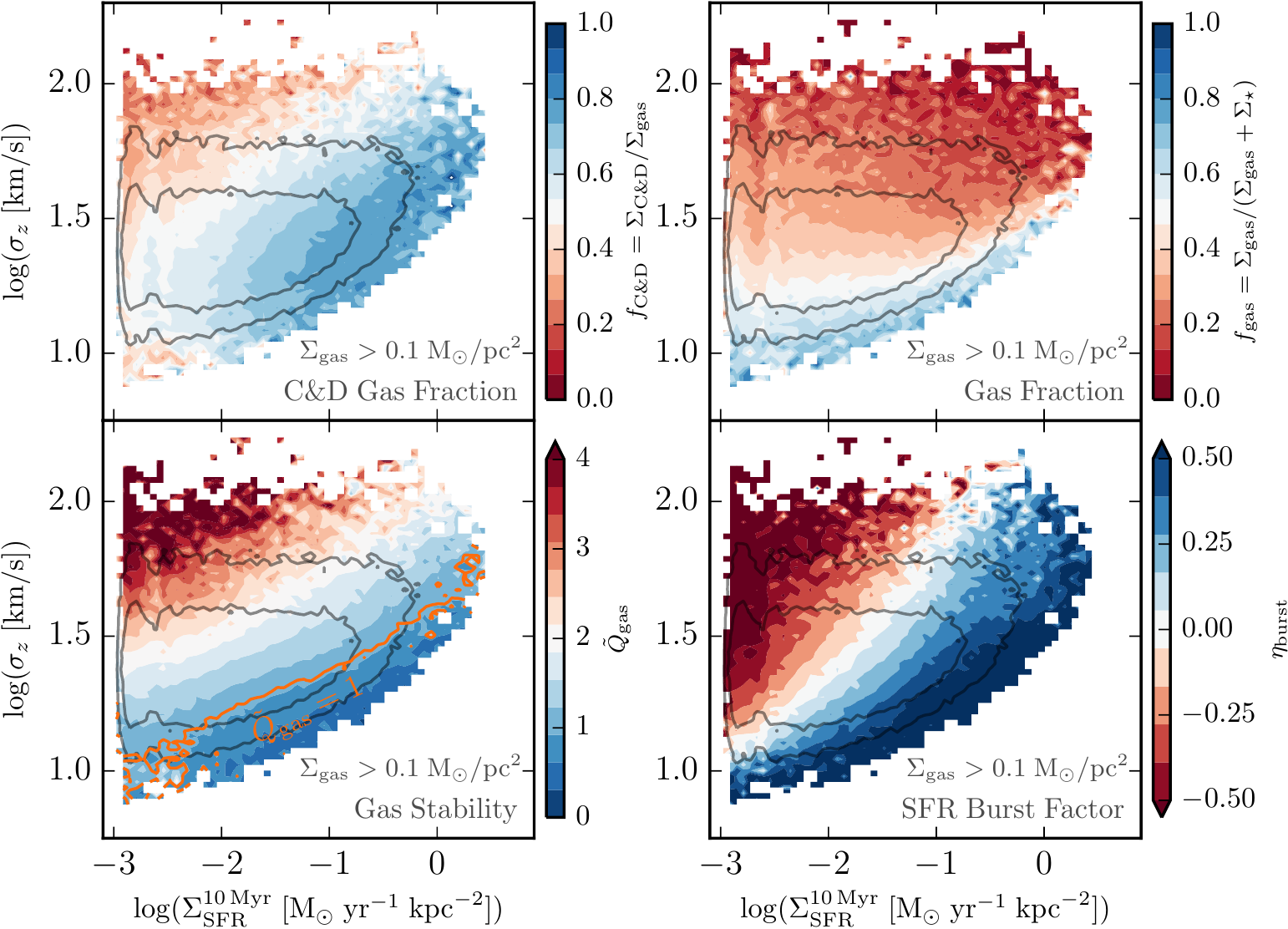}
	\caption{Distribution of $\sigma_z$ and SFR (as Fig.~\ref{fig:tracers}, grey contours show 75, 95\% data inclusion regions) colored by various gas and star formation properties.
	\textbf{Top Left:} Cold and dense gas fraction ($f_{\rm C\&D} = \Sigma_{\rm C\&D}/\Sigma_{\rm gas}$ in pixels). C\&D gas fraction rises as SFR increases at constant velocity dispersion. 
	\textbf{Top Right:} Gas fraction  ($f_{\rm gas} = \Sigma_{\rm gas}/(\Sigma_{\rm gas}+\Sigma_\star)$ in pixels). As $\Sigma_{\rm SFR}$ scales with $\Sigma_{\rm gas}$, at fixed $\Sigma_{\rm SFR}$ lower $f_{\rm gas}$ is driven by higher $\Sigma_\star$ and $\Sigma_{\rm tot}$
	Gas-rich regions have the lowest velocity dispersions for their SFRs. For a constant amount of gas/star formation, gas in a deeper potential (higher stellar surface densities) must have larger velocity dispersions to maintain stability/equilibrium.  
	\textbf{Bottom Left:} Turbulent Toomre-Q parameter $\tilde Q_{\rm gas}$ (Eq.~\ref{eq:Q}).  Orange line denotes $\tilde Q_{\rm gas} = 1$ threshold for instabilities (i.e., pixels in this region of $\Sigma_{\rm SFR}^{\rm 10 \, Myr}-\sigma_z$ space are on-average unstable).  Less stable gas (at fixed $\sigma_z$) produces higher 10 Myr average star formation rates.
	\textbf{Bottom Right:} Star formation ``burstiness'', $\eta_{\rm burst}$.  Regions with low ratios of past to current SFRs have low $\sigma_z$ at constant 10 Myr-averaged SFR, corresponding with regions of $\tilde Q_{\rm gas} \lesssim 1$.}
	\label{fig:gasprops}
\end{figure*}
Figure~\ref{fig:gasprops} shows the $\sigma_z$--SFR relation colored by average pixel gas and star formation properties.  The top left panel is colored by the mass fraction of cold and dense gas in the neutral (atomic + molecular) ISM.  This fraction is a rough proxy for the molecular gas fraction on $\sim$kpc scales.  Here we clearly see that for a given amount of turbulence (velocity dispersion) more molecularly rich gas has higher star formation rates on average.  It is not surprising that higher dense gas fractions correspond with higher star formation rates, considering that star formation occurs predominantly in cold and dense gas in molecular clouds.

However, this correlation between high dense gas fractions and star formation rates may appear as a bias in the observations towards higher star formation rates at constant $\sigma$ in the $\sigma$--$\Sigma_{\rm SFR}$ plane, when correlating star formation rates with velocity dispersions that are pegged dense gas regions alone (e.g., those traced by CO $J = 1 \rightarrow 0$ or higher transition emission).  Given the difficulty/expense in measuring atomic gas (H{\scriptsize I}) on small ($\lesssim 100$ pc) scales, a full accounting of the turbulent momentum in both the atomic and molecular ISM on kpc scales may not necessarily be done, tending towards over-weighting the dynamics of the densest gas, resulting in a `steep' $\sigma_z$--SFR relaiton.  

The top right panel shows the overall gas fraction, $f_{\rm gas} = \Sigma_{\rm gas}/(\Sigma_{\rm gas} + \Sigma_\star)$, another factor that clearly affects the normalization of the velocity dispersion for a given star formation rate.  We see there is little variation in average gas fractions for a given amount of turbulence (velocity dispersion). Instead, at a given star formation rate, lower gas fractions yield larger turbulent velocity dispersions across 3 dex in SFRs.  At lower gas fractions, for a given amount of gas (presumably, here SFRs still correlate with gas surface densities as seen before in the FIRE-1 simulations by \citealt{Orr2018}), a larger stellar component in the disk produces a deeper potential well for the gas to stabilize itself in.  Just as larger $\Sigma_{\rm gas}$ requires higher $\sigma_z$ in Eq.~\ref{eq:Q}, so does larger $\Sigma_\star$.

As neutral (atomic + molecular) gas fractions approach zero at the highest velocity dispersions ($\sigma_z > 100$ km/s), it is evident in these galaxies that an appreciable fraction of the gas is becoming ionized and is in a hot, usually outflowing state.  That gas ought to no longer be considered either neutral in nature, nor in any equilibrium state (with respect to dynamics, e.g., hydrostatic/turbulent support, or related to its current star formation rate).

The bottom left panel of Figure~\ref{fig:gasprops} shows the velocity dispersion--SFR relation colored by the average gas stability (our modified Toomre-Q, Eq.~\ref{eq:Q}).  The average trend is similar to the trend in cold and dense gas fraction and star formation rates: less stable gas is both more predominantly cold and dense in nature, and has higher star formation rates for a given amount of turbulence (velocity dispersion).  Interestingly, the trend follows a $\sigma \propto \Sigma_{\rm SFR}^{\sim 1/6}$ relation for constant $\tilde Q_{\rm gas}$ (see the orange line for $\tilde Q_{\rm gas} =1$, with its $\sim$1/6 slope).  This power law slope is shallower than expected for a feedback regulated, turbulent star formation environment, as derived in Eq.~\ref{eq:sigmaFB}, suggesting that a simple turbulently regulated feedback framework alone is insufficient to describe the observed $\sigma$-$\Sigma_{\rm SFR}$ relation, and processes that raise $\sigma$ at low SFR (e.g., non-negligible thermal support) or depress it at high SFR (e.g., momentum going into outflows or a hot ISM phase instead of neutral gas $\sigma$) are required at a minimum. 

That there are patches at (albeit) low 10 Myr-averaged SFR that have such high values of $\tilde Q_{\rm gas}$($\gtrsim 3$) is interesting: regions with $\tilde Q_{\rm gas}>1$ are expected to be stable against star formation, i.e., to have no/very little current (instantaneous) star formation.  The fact that high $\tilde Q_{\rm gas}$ regions do have non-zero 10 Myr star formation rates suggests that star forming regions may be able to regulate and inject sufficient stabilizing feedback momentum on sub-10 Myr timescales. Indeed, this timescale corresponds to both the bulk of ionizing radiation from massive stars and the first (few) SNe in a GMC.

As well, all of the velocity dispersions calculated for the neutral gas are in excess of $\sim 10$ km/s, above the sound speed for 8000 K atomic gas $c_s \approx 6$ km/s.  These in excess of $10$ km/s velocity dispersions are in line with many observations of the star-forming ISM, where few, if any, star-forming regions are purely thermally supported even at the lowest star formation rate surface densities \citep{Stilp2013, Stilp2013a}.

\subsection{Velocity Dispersions and SFR Timescales}
The bottom right panel of Figure~\ref{fig:gasprops} investigates correlations within the $\sigma_z$--SFR relation, when considering the 10 Myr average star formation rate and neutral gas $\sigma_z$, on the recent star formation history of the pixels.  Here we color the relation by the star formation burst measure $\eta_{\rm burst}$, where bluer (redder) patches indicate rising (falling) SFRs in time, corresponding with the injection of feedback over time as patches evolve from blue to red following a starburst. 

Probing star formation rates on 100 Myr vs. 10 Myr timescales provides an interesting test of the feedback-regulation picture, where longer star formation rate tracers are sensitive to regions that have already injected all, or a significant fraction, of their feedback budget.  Thus, shorter timescale tracers are indicative of the current demands of the ISM, feedback-wise, whereas longer tracers should trace the more-averaged history of the momentum balance in the ISM.  Hence, bluer patches indicate currently unstable star-forming regions, while redder regions reflect those that have presumably received the bulk of feedback from young stars following a starburst. The lower envelope of the relation appears to follow a curve where a single star formation event has occurred in the last 10 Myr, such that $\Sigma_{\rm SFR}^{\rm 100 \, Myr} \sim \Sigma_{\rm SFR}^{\rm 10 \, Myr}/10$.  And so, at low recent (10 Myr) star formation rates, the amount of star formation that has occurred in the past 100 Myr is strongly correlated with the amount of turbulence (velocity dispersion) in the ISM.  There is essentially space between the velocity dispersion floor and ceiling (cf. \S~\ref{subsec:outflows}) for on-off star formation cycles to occur over 100 Myr timescales.  However, at high recent star formation rates, both the dynamical times of the regions become shorter (thus leaving little room temporally for on-off star formation modes) and the velocity dispersions cannot be driven much higher before hitting an outflow threshold.  Thus, the 100 Myr-averaged tracer at very high recent star formation rates approaches $\Sigma_{\rm SFR}^{\rm 100 \, Myr} \approx \Sigma_{\rm SFR}^{\rm 10 \, Myr}/10$ (modulo stellar evolution mass loss correction factors) with a single star formation event. At these high SFRs, the ISM effectively has no feedback history locally, and is in an unstable, highly star-forming state.

\subsection{Depletion Times in Neutral and Cold \& Dense Gas}
\begin{figure*}
	\centering
	\includegraphics[width=0.92\textwidth]{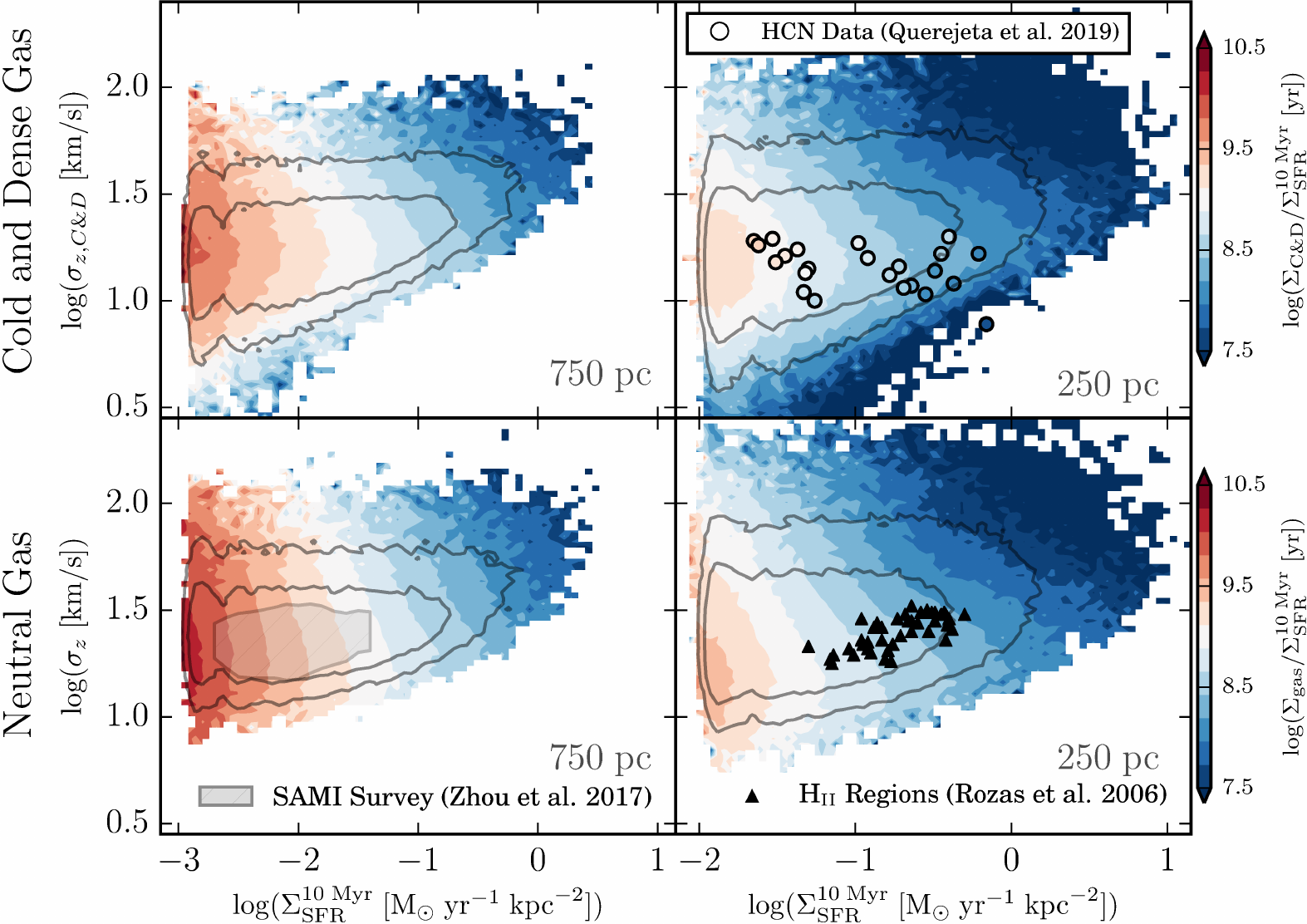}
	\caption{We compare $\sigma_z$ (as Fig.~\ref{fig:tracers}) in cold and dense gas (\emph{top row}) or all neutral gas (\emph{bottom row}) with 750 pc (\emph{left column}) or 250 pc (\emph{right column}) pixels, colored by the gas depletion time, to corresponding observations \emph{also colored} (where possible) by depletion time. Dispersions ($\sigma_z$) are only weakly correlated with depletion time at fixed $\Sigma_{\rm SFR}$, as observed. Cold and dense gas exhibit two regimes of short depletion times, gas with high $\sigma_z$ and SFRs, and low $\sigma_z$ at fixed SFR tend towards short ($\sim$few 10$^7$ yr) dense gas depletion times.   The core of the simulation data distributions, and observations agree well.}
	\label{fig:Tdep}
\end{figure*}
Figure~\ref{fig:Tdep} explores how gas depletion times vary across the velocity dispersion--SFR relation for both 250 pc and 750 pc pixel sizes.  We explore both how the depletion time of the cold and dense gas and of all neutral gas vary with their respective velocity dispersions for the 10 Myr-averaged tracer of star formation.  Here we directly compare with three observational datasets: two H{\scriptsize II} velocity dispersions and H$\alpha$ SFR studies \citep{Rozas2006, Zhou2017}, and one of the velocity dispersions and depletion times in HCN (tracing very cold and dense gas, \citealt{Querejeta2019}).  Neither the \citet{Rozas2006} nor the \citet{Zhou2017} datasets include gas depletion time estimates, and so only constrain the extent of our $\sigma$-SFR relation.  With these, we find good agreement.  

The \citet{Querejeta2019} data, however, did include measurements of the depletion time of the dense gas.  We compare our cold and dense depletion time data directly against the $\Sigma_{\rm SFR}/\Sigma_{\rm HCN}$ with the same colorbar, and find very good agreement. 
Particularly interesting for comparison with the \citet{Querejeta2019} data, there are two distinct (bluer) regions with shorter dense gas depletion times on 250 pc scales: (1) low velocity dispersions for a given 10 Myr-averaged SFR, which is the scaling that their observational dataset covers, and (2) high 10 Myr-averaged SFRs \emph{and} high $\gtrsim 30$ km/s dense gas velocity dispersions, which their dataset did not trace.  At 750 pc pixel size, the short depletion time region at low velocity dispersions is not seen as strongly, with $\tau_{\rm dep} = \Sigma_{\rm gas}/\Sigma_{\rm SFR}$ falling to $\sim 10^{8.5}$ yr at low velocity dispersions.

Interpreting further the case where dense gas depletion times become short at low $\sigma_{z,C\&D}$, this appears to happen in pixels in the simulations where only a small fraction of the ISM is cold and dense (and its contribution to the overall turbulent momentum is low).  This case occurs in galactic outskirts where the ISM is considerably more diffuse, but requires some stabilizing feedback.  There, the ISM evolves to produce just enough dense gas to form stars, but in the process is rapidly consumed.  

\subsection{Outflow-prone Gas Fractions: Balancing Self-gravity and Integrated Feedback}\label{subsec:outflows}
\begin{figure*}
	\centering
	\includegraphics[width=0.97\textwidth]{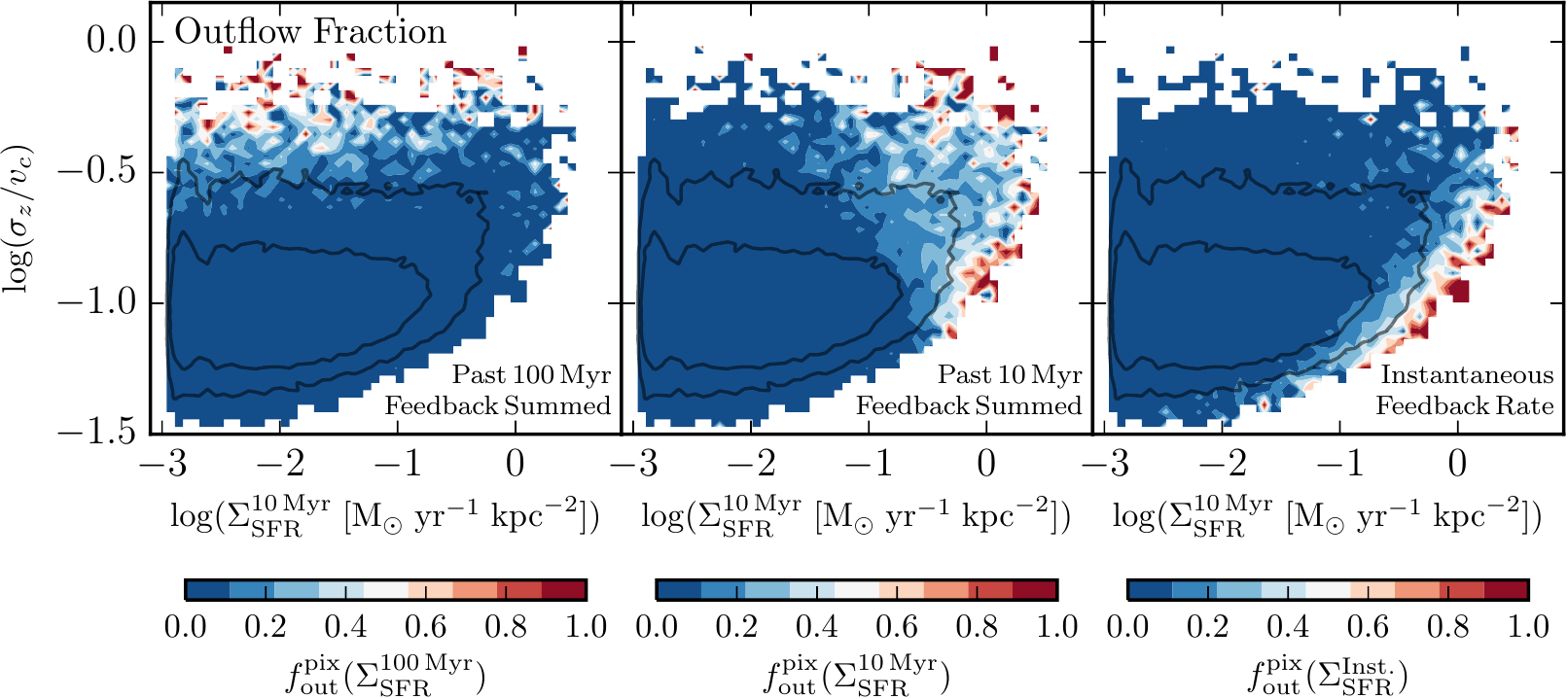}
	\caption{Distribution of $\Sigma_{\rm SFR}^{\rm 10\, Myr}$ versus $\sigma_z$ relative to the local circular velocity $v_c$ (style as Fig.~\ref{fig:gasprops}), now colored by the theoretical prediction for whether the gas $f_{\rm out}$ should be unbound by feedback in ``steady state'' (see \S~\ref{subsec:outflows} in text) from the model in \citet{Hayward2017} (given $\Sigma_{\rm SFR}$, $v_c$, and $R$).  Coloring is by the average fraction of pixels whose gas meets the outflow criterion at that point in $\Sigma_{\rm SFR}$--$\sigma_z/v_c$ space.  If we calculate $f_{\rm out}$ from the SFR averaged on 100 Myr timescales (\emph{left panel}), i.e., $\sim t_{\rm dyn}$ of the galaxy it neatly matches where gas is outflowing ($\sigma_z \gtrsim v_c$); but if we calculate $f_{\rm out}$ using the ``instantaneous'' SFR (\emph{right panel}), it corresponds not to where present outflows exist (the outflows have not developed), but it defines the ``upper limit'' of $\Sigma_{\rm SFR}$, where local ``early'' feedback shuts down dense gas collapse.  Averaging over 10 Myr timescales (\emph{middle panel}), there is an intermediate scenario, where outflows are developing and gas is being driven to high $\sigma_z$ approaching local escape velocities.}
	\label{fig:outflows}
\end{figure*}

Figure~\ref{fig:outflows} explores the role of an outflow threshold in setting the limits of the turbulence in the ISM for a given star formation rate and recent star formation history.   In previous sections, we explored how the velocity dispersion--SFR relation depends on properties of the gas and, to an extent, recent star formation.  However, the `edges' in the relation are correlated seemingly with the (neutral) gas fractions (high- and low-$\sigma_z$ limit) and gas stability ($\tilde Q_{\rm gas}$, lower $\sigma_z$ limit at constant SFR).  Naturally, regions that approach having no neutral gas ($f_{\rm gas} \rightarrow 0$) will no longer be able to support star formation.  As well, patches of the galaxy can only be so unstable ($\tilde Q_{\rm gas} \rightarrow 0$).  The third edge in the relation here, that of a minimum star formation rate, is resolution limited in these simulations, and not indicative of a physical cutoff in star formation rates.  However, there was not a clear reason physically for either the $f_{\rm gas} \rightarrow 0$ or the $\tilde Q_g \rightarrow 0$ edge to occur where they did, in terms of the normalization of $\sigma_z$.  

Rescaling the velocity dispersions by their ratios to the local circular velocity in the galaxy, as is often discussed when considering galactic outflows, we see firstly that the vast majority ($> 75$ \% by area) of the galaxies have disk aspect ratios $v_c/\sigma_z$ between $\sim 4-20$.  The simulations have, by and large, thin disks over most of their areas. 

We consider an outflows-threshold from of \citet{Hayward2017}, where there is a surface density threshold in the gas for which a given amount of feedback per area in time $(P/m_\star)\Sigma_{\rm SFR}$ can accelerate that patch of gas to the local escape velocity in a coherence time (roughly equivalent to an eddy crossing time).  This threshold, Eq. 5 in \citet{Hayward2017}, is $\Sigma_{\rm gas} < (P/m_\star) \Sigma_{\rm SFR} / \sqrt{2}R\Omega^2$.  In Figure~\ref{fig:outflows}, we shade the ($\sigma_z/v_c$)-$\Sigma_{\rm SFR}$ relation by the fraction of pixels that have gas surface densities below this threshold and thus find themselves `outflow-prone' at that position in ($\sigma_z/v_c$)-$\Sigma_{\rm SFR}$ space.  Interestingly, the different edges of the ($\sigma_z/v_c$)-$\Sigma_{\rm SFR}$ relation for neutral gas velocity dispersions and the 10 Myr star formation rate tracer, excluding the mass resolution limit, map to the regimes where significant fractions of gas become outflow-prone depending on the different timescale star formation tracers.  As the local coherence time is on the order of the eddy crossing time, all of the star formation tracer timescales explored are roughly equivalent or shorter than $\sim 100$ Myr, so each tracer integrates parts of the feedback that end up driving outflows (i.e., the instantaneous tracer only tracing radiative pressure and winds, whereas the 10 Myr-averaged tracer integrates both those and early Type-II SNe).  

Focusing on the material that is outflow-prone when considering the instantaneous SFR, it appears that the hard-limit in the velocity dispersion-SFR relation at constant $\sigma$ is set by the amount of momentum that the ISM can absorb before appreciable fractions of it are blown out efficiently.  On short ($< 10$ Myr) timescales, as per the discussion in \S~\ref{sec:theory1}, it is reasonable to consider $\sigma$ constant or only very slowly varying as star formation \emph{fires} up.  On the other extreme of our star formation tracer timescales, when we consider the outflow fractions as calculated by the 100 Myr tracer, we see that the outflow-prone material traces the most extreme $\sigma_z/v_c$, above $\log(\sigma_z/v_c) \geq -0.5$.  That this gas has the highest amount of dispersion support ($\propto \sigma_z/v_c$) is understandable, as on 100 Myr timescales, feedback is able to pump velocity dispersions on disk scales, and regions that are marginally outflow-prone and still have any gas left will have the highest velocity dispersions.  After all, these kpc-patches have absorbed the full feedback from one or more young star clusters.  The outflow fraction approaching unity for 100 Myr timescale star formation rates in disk environments thus provides a physical reason for the top edge of the ($\sigma_z/v_c$)-$\Sigma_{\rm SFR}$ relation.  

The 10 Myr SFR tracer prediction, lying between these two natural extreme timescales, connects the two limits, representing regions that are seemingly either maximally forming stars locally (and regulated by prompt feedback like radiation pressure and stellar winds), or have velocity dispersions approaching escape velocity, presumably due to recently injected feedback.

\subsection{Gas Properties Correlating with Gas Stability Budgets: Does the Degree of Turbulence Matter?}\label{subsec:QTdep}
\begin{figure*}
	\centering
	\includegraphics[width=0.95\textwidth]{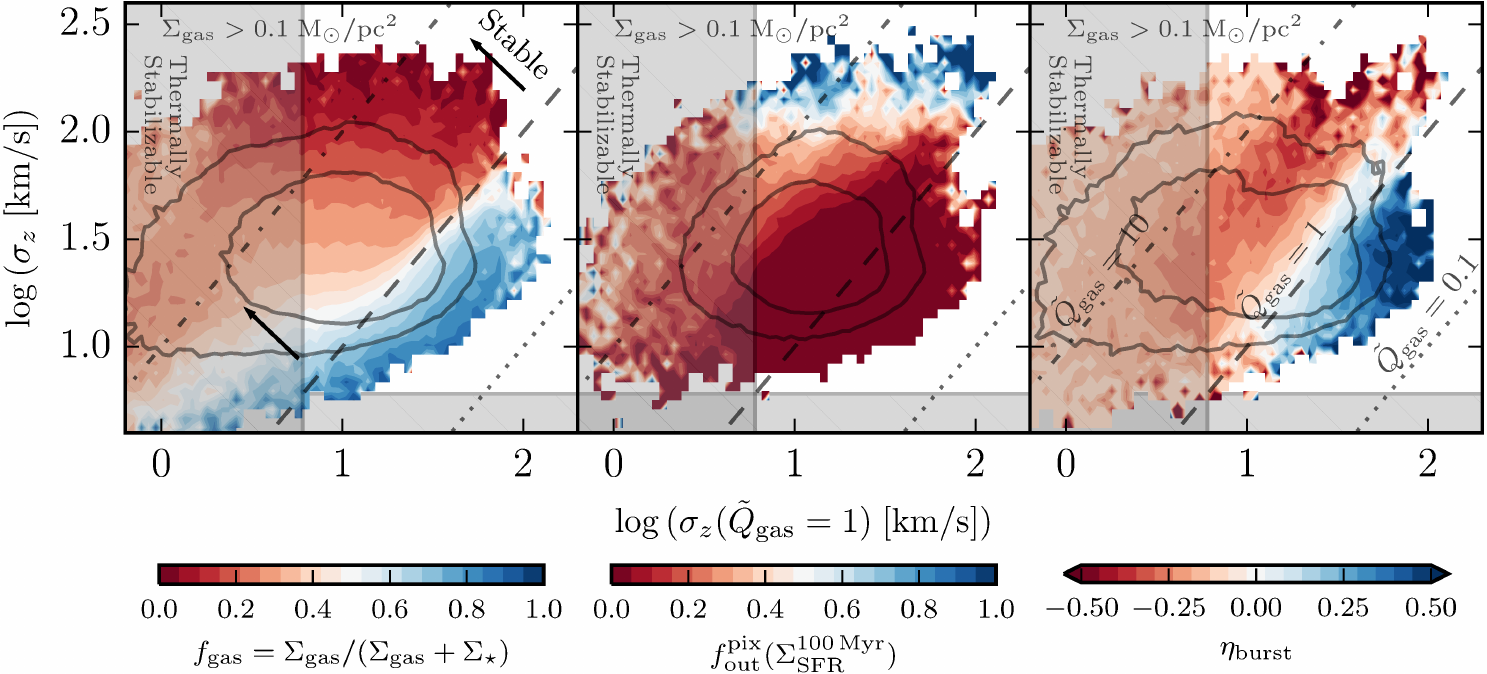}
	\caption{Distributions of gas velocity dispersion required for a given pixel (750 pc) to have $\tilde Q_{\rm gas}=1$ (Eq.~\ref{eq:Q}), versus actual $\sigma_z$ in neutral gas, colored by average pixel gas fraction (\emph{left}), whether the 100 Myr SFR predicts outflows (\emph{middle}), or star formation ``burstiness'' $\eta_{\rm burst}$ (\emph{right}). Considering regions with $\sigma_z(\tilde Q_{\rm gas}=1) > 6$ km/s (i.e., regions that \emph{cannot} be only thermally supported), gas rich regions ($f_{\rm gas} \rightarrow 1$) are unstable to gravitational collapse and fragmentation ($\tilde Q_{\rm gas} < 1$), whereas the most gas poor regions ($f_{\rm gas} \rightarrow 0$) have high velocity dispersions and lie in the $1 < \tilde Q_{\rm gas} \lesssim 10$ regime.  Similarly, high $\sigma_z$ is a stronger predictor of whether or not a region is likely to host outflows ($f_{\rm out}^{\rm pix} \rightarrow 1$) than $\tilde Q_{\rm gas}$, provided $\tilde Q_{\rm gas} >1$.  Interestingly, for regions that \emph{can} be thermally stabilized, the fraction of regions capable of hosting outflows rises strongly with $\tilde Q_{\rm gas} \gtrsim 3$. Regions with rapidly rising (decaying) local SFRs correspond to regions with $\tilde Q_{\rm gas} <1$ ($\tilde Q_{\rm gas} >1$).}
	\label{fig:WVplots}
\end{figure*}

\begin{figure}
	\centering
	\includegraphics[width=0.49\textwidth]{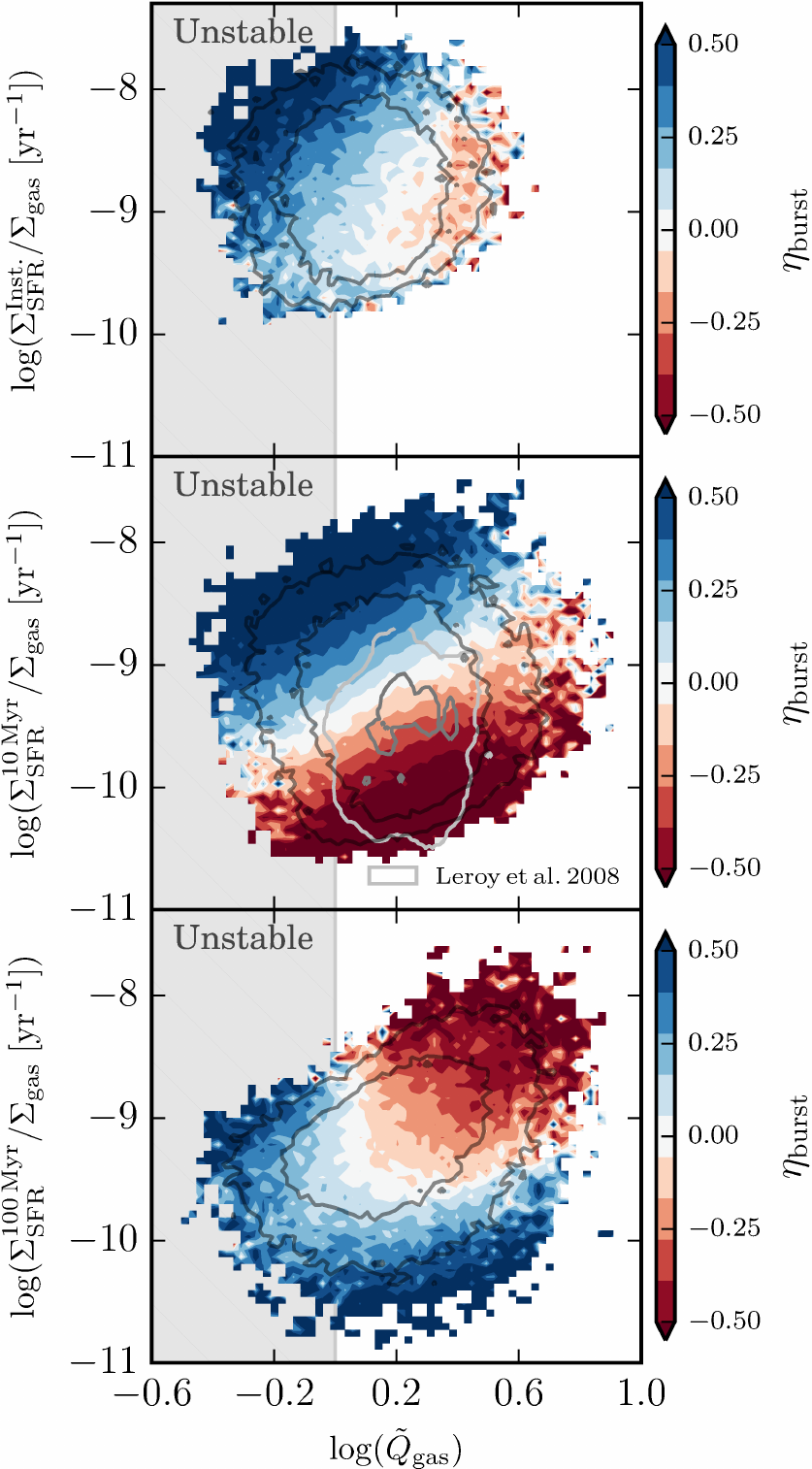}
	\caption{Distribution of pixel inverse depletion times (using SFR averaged on different timescales), labeled vs. local $\tilde Q_{\rm gas}$, colored by star formation ``burstiness'' parameter $\eta_{\rm burst}$ (as Fig.~\ref{fig:galsm12b-i}), compared where appropriate with spatially resolved observations of \citet{Leroy2008}.  Gray contours indicate 95\% and 75\% data inclusion regions (by pixel number).  Pixels are cut for $\Sigma_{\rm gas} < 0.1$ M$_\odot$ pc$^{-2}$ and outside of $3 < R/{\rm kpc} < 10$.  As gas clusters very close to $\tilde Q_{\rm gas}=1$, there is almost no direct correlation between $\tilde Q_{\rm gas}$ and (inverse) depletion time.  Gas with rising (decaying) SFRs does tend towards smaller (larger) $\tilde Q_{\rm gas}$. }
	\label{fig:Q-tdep}
\end{figure}
Figure~\ref{fig:WVplots} plots the predicted velocity dispersions in each pixel, assuming $\tilde Q_{\rm gas} = 1$ versus the measured $\sigma_z$, and colors the relation by average gas fractions, outflow-prone fractions (calculated with the 100 Myr tracer) and $\eta_{\rm burst} = (\Sigma_{\rm SFR}^{\rm 10 \, Myr}-\Sigma_{\rm SFR}^{\rm 100 \, Myr})/(\Sigma_{\rm SFR}^{\rm 10 \, Myr}+\Sigma_{\rm SFR}^{\rm 100\, Myr})$, a measure of the star formation bursty-ness.  If all the patches of the galaxies had $\tilde Q_{\rm gas} = 1$, then the whole dataset would fall on the dotted $\tilde Q_{\rm gas} = 1$ line, so the plots show how these quantities vary with how far from equilibrium individual regions are.  The floor in $\sigma_z$ appears where the atomic phase of the ISM (stable at $T \approx 6000$ K) provides non-negligible thermal support with $c_s \approx 6$ km/s.  The required $\sigma_z$ to stabilize the gas (for $\tilde Q_{\rm gas} = 1$) can fall to arbitrarily low values, and any amount of thermal support may be sufficient to maintain stability (also represented by a greyed-out patch for $\sigma_z(\tilde Q_{\rm gas} = 1) < 6$ km/s).

Concerning local gas fractions $f_{\rm gas}$ In the simulations, regions with $\tilde Q_{\rm gas} < 1$ are on-average gas rich with $f_{\rm gas} > 0.5$.  The most stable regions requiring turbulent support $\sigma_z(\tilde Q_{\rm gas} = 1) > c_s \approx 6$ km/s are the least gas rich, with higher values of $\sigma_z$ having $f_{\rm gas} \rightarrow 0$.

The middle panel shows the outflow-prone fraction calculated using the 100 Myr average star formation rate.  High velocity dispersion, low gas fraction regions are the only areas with significant outflow-prone fractions.  This suggests that significant feedback events over 100 Myr timescales have had a chance to drive these dispersions to high values, i.e.,$\tilde Q_{\rm gas} \gg 1$.  Evidently, these regions do not just have high velocity dispersions because they have deep disk potentials with high $\Sigma_\star$, though these are correlated to an extent.

Finally, looking at an indicator of how bursty star formation is, i.e., the relative amount of star formation occurring now versus in the recent past, $\eta_{\rm burst}$, we see that the $\tilde Q_{\rm gas} = 1$ threshold is a demarcating line between an abundance of current star formation as compared to the past (bluer shading) and vice versa.  The right panel of Figure~\ref{fig:WVplots} shows that vigorous star formation occurs in the ISM when the gas is gravitationally unstable against fragmentation, and collapses on the Toomre-scale (here, the disk scale height).

Exploring this last aspect further, Figure~\ref{fig:Q-tdep} shows how the gas depletion time (calculated by the time for neutral gas consumption with two different star formation tracers, the gas instantaneous and 10 Myr average star formation rate) correlates with gas stability using  $\tilde Q_{\rm gas}$.  The relation is shaded by the star formation burst indicator $\eta_{\rm burst}$.  Importantly, we exclude gas outside 3 kpc $< R <$ 10 kpc, and with $\Sigma_{\rm gas} < 0.1$ M$_\odot$ pc$^{-2}$, as it is either not shielded and likely has significant thermal support, or is at radii with a rising rotation curve and has significant shear across the 750 pc pixels, such that the assumption of a flat rotation curve in calculating $\tilde Q_{\rm gas}$ breaks down and it is no longer a good measure of gas stability.

When considering the instantaneous star formation rate gas depletion time, we see that $\tilde Q_{\rm gas}$ alone is a good predictor of $\eta_{\rm burst}$, with less stable regions having significant current relative to past star formation.  Notably, there are few regions with $\eta_{\rm burst}  \lesssim -0.5$, i.e.,pixels that have significant instantaneous (current) star formation and have recent star formation histories dominated by past ($10$ Myr$ < t_{\rm ago} < 100$ Myr) star formation.

On the other hand, the 10 Myr average star formation rate gas depletion time panel shows two regimes: (1) short depletion times are dominated by regions with more recent versus past star formation ($\eta_{\rm burst} > 0.5$) and are on-average less stable ($\tilde Q_{\rm gas}$ is $\sim 0.3$ dex lower), and (2) long-depletion time regions have slightly higher values of $\tilde Q_{\rm gas}$ on-average, and are dominated by past versus recent star formation.  The on-off cycle picture is further borne out when considering the 100 Myr-averaged gas depletion times (bottom panel).  The patches with the shortest depletion times averaged over 100 Myr (i.e., most integrated star formation per unit \emph{remaining} gas) are the most stable on average.  That there exist bursts of star formation agrees with recent work by \citet{Orr2019}, where on-off cycles of star formation can be driven on $1/\Omega$ timescales by feedback from Type II SNe and their $\sim 40$ Myr delay time distribution.
\section{Discussion}
\subsection{A Hierarchy of Timescales: Why is $\sigma_z$-$\Sigma_{\rm SFR}$ so flat?}\label{sec:theory1}
It is notable that for three dex in (10 Myr-averaged) star formation rates, $\sigma_z$ hardly budges (cf. Figure~\ref{fig:tracers}), the lower envelope in velocity dispersions notwithstanding.  This flatness in velocity dispersions however, like the discussion regarding scatter in the resolved Kennicutt-Schmidt relation at high spatial resolution \citep{Schruba2010,Kruijssen2014,Orr2019}, may be understood through a discussion of a hierarchy of timescales.  Within the framework of supersonic turbulence dominating the velocity dispersions, the turbulent momentum in a patch of the ISM is decaying on an eddy-crossing time $t_{\rm eddy} \sim 2/\Omega$, and so $\dot\sigma_z \sim -2\sigma_z\Omega$.  For the Milky Way-like galaxies explored here, $\Omega = v_c/R \sim$ 250 km/s /10 kpc $\sim 25$ Gyr$^{-1}$; thus, $t_{\rm eddy} \sim 80$ Myr.  This timescale is far longer than the free-fall time for a GMC with a mean density of $n=100$ cm$^{-3}$, $t_{\rm ff} = \sqrt{3 \pi / 32 G m_p n} \approx 5$ Myr (see also the simulated GMC lifetimes of \citealt{Grudic2018}).  As a result, while $\sigma_z$ only slowly evolves as turbulence is dissipating, short timescale tracers of star formation do not persist long enough to trace the effects of feedback as it is felt/absorbed by the ISM on the relevant scales for the gas velocity dispersions (of course, `prompt' feedback like winds and ionizing radiation are locally felt immediately).  Considering that the feedback momentum from supernovae in a star cluster is meted out over a period of $\sim 5-40$ Myr (to say, $t_{\rm fb} = 40$ Myr), it is understandable that some evolution in $\sigma_z$ is seen for $\sigma$-$\Sigma_{\rm SFR}$ when considering the 100 Myr-averaged star formation rate tracer. This longer-timescale star formation rate tracer is able to overlap with the entire feedback injection period, not just the start of it.  This differential-timescales picture is especially true when considering the gas instantaneous star formation rate, where there is effectively no evolution in the velocity dispersions as the star formation rate ramps up and down per the local ISM conditions.  Instantaneous star formation rates do not appear to have any correlation with velocity dispersions in the ISM (barring the slight lower envelope in velocity dispersions for a given instantaneous SFR, seen more strongly for the 10 and 100 Myr SFR relations).  The initial distribution of velocity dispersions is simply smeared out across the range of star formation rates as they rapidly rise and fall.  And so, it is expected that we see a flat distribution in velocity dispersion, and a slightly positive slope in it, for 10 Myr and 100 Myr tracers of star formation, respectively, with a hierarchy of timescales: $t_{\rm ff} \ll 10\; {\rm Myr} < t_{\rm fb} < 100 \; {\rm Myr} \sim t_{\rm eddy} \sim 1/\Omega$.

\subsection{What Drives Velocity Dispersions (Theoretically)?} \label{subsec:scalings}
\begin{figure*}
	\centering
	\includegraphics[width=0.97\textwidth]{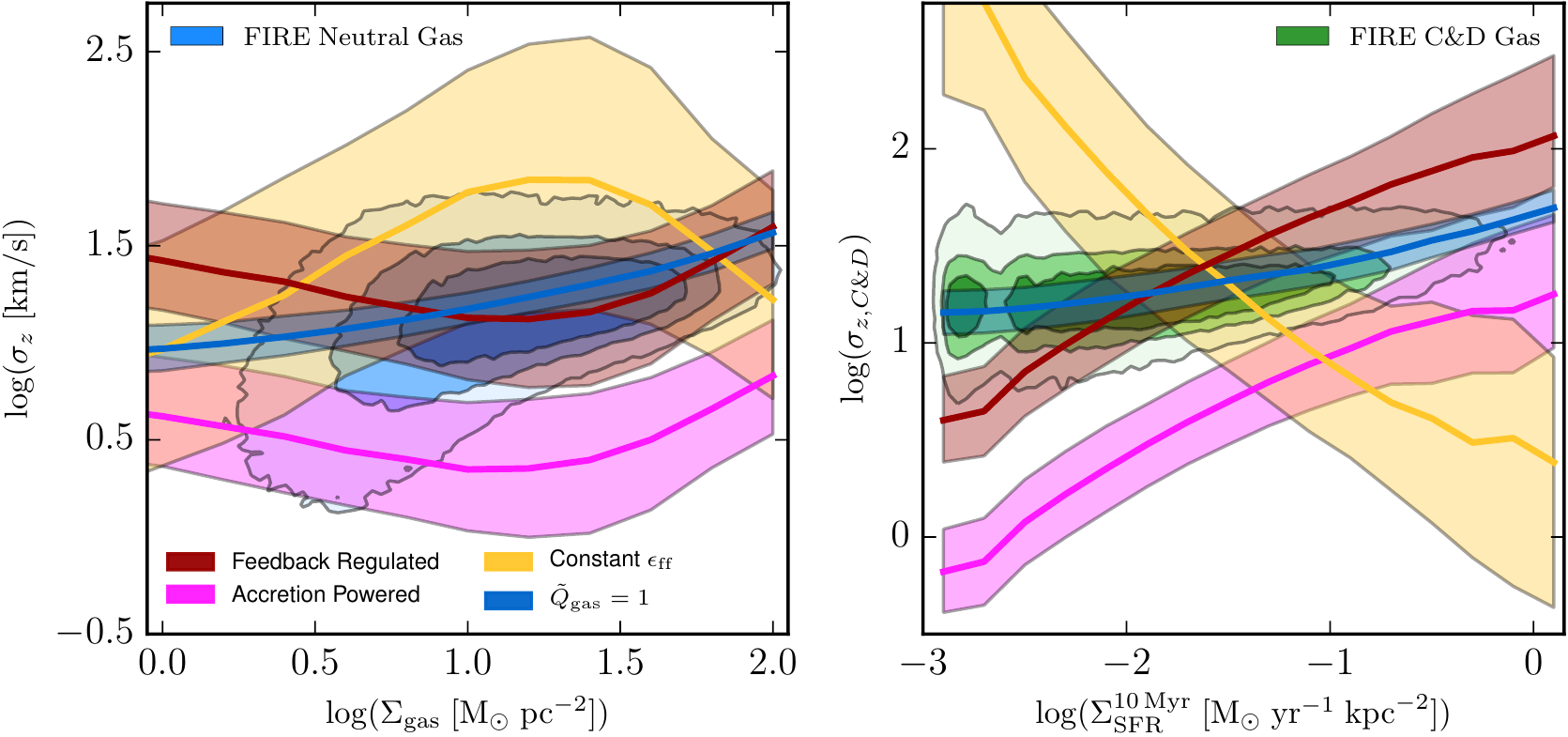}
	\caption{Pixel by pixel predictions of velocity dispersions by ``Feedback-regulated'' model (cardinal, \S~\ref{subsec:scalings} \ref{subsec:FBtheory}), a ``Constant-efficiency per free-fall time'' model (gold, \S~\ref{subsec:scalings} \ref{subsec:CEtheory}), assuming $\tilde Q_{\rm gas} =1$ (blue, Eq.~\ref{eq:Q}), and an ``Accretion-powered turbulence'' model (magenta, \S~\ref{subsec:scalings} \ref{subsec:APtheory}) compared against actual dispersions in the FIRE simulations (background blue and green shaded contours, style as Fig.~\ref{fig:tracers}).  In both panels, model shaded regions denote interquartile region (with median as solid line), in bins of $\Sigma_{\rm gas}$ and $\Sigma_{\rm SFR}^{\rm 10 \, Myr}$. All scalings are calculated and presented with the 10 Myr average star formation rate tracer.
	\textbf{Left:}  Predicted $\sigma_z$ for star formation models and $\tilde Q_{\rm gas} =1$ versus measured neutral gas $\sigma_{z}$ as a function of $\Sigma_{\rm gas}$.  The Feedback-regulated and $\tilde Q_{\rm gas} =1$ models agree well with the core of the measured gas velocity dispersions, whereas the constant-efficiency and accretion-powered models over- and under-prediction velocity dispersions in the gas, respectively. 
	\textbf{Right:} Velocity dispersion-SFR relation, showing both model distributions and the FIRE dataset (green shaded contours). Here the different slope scalings of the star formation models with respect to $\Sigma_{\rm SFR}$ are on display: feedback-regulation predicts rising $\sigma$ with $\Sigma_{\rm SFR}$, whereas constant-efficiency star formation predicts falling dispersions, and is in strong disagreement with the simulations. $\tilde Q_{\rm gas} =1$ falls on top of the FIRE data at all SFRs.  $\tilde Q_{\rm gas} =1$ is a closer predictor than either at all $\sigma_z$.}
	\label{fig:FBvG}
\end{figure*}

\begin{figure}
	\centering
	\includegraphics[width=0.47\textwidth]{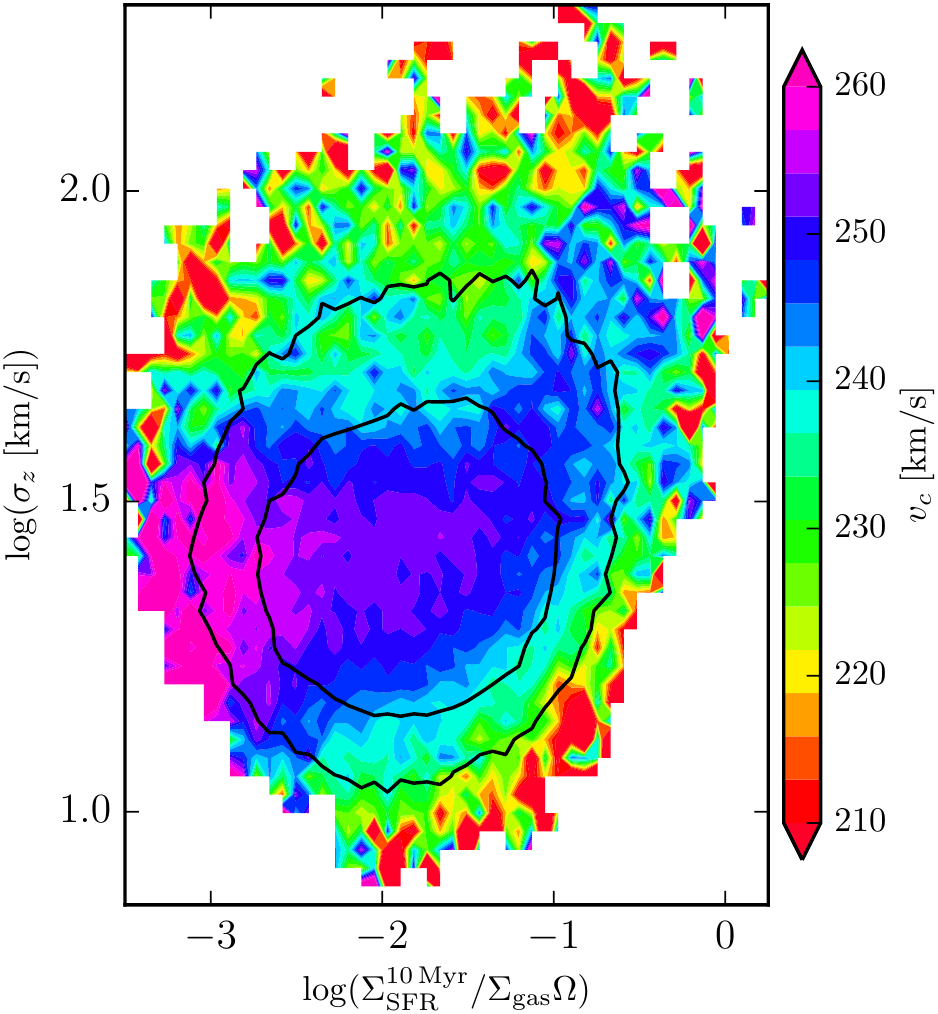}
	\caption{Neutral gas velocity dispersion (in 750 pc pixels) as a function of star formation efficiency in the FIRE simulations, colored by average circular velocity ($v_c$) of pixels.  Black contours indicate 75\% and 95\% data inclusion regions.  Both the feedback-regulated (Eq.~\ref{eq:sigmaFB}) and accretion powered (Eq.~\ref{eq:sigmaAP}) models predict $\sigma \propto \Sigma_{\rm SFR}/\Sigma_{\rm gas}\Omega$.  However, the accretion powered model predicts an additional linear dependence on $v_c$ (the feedback-regulated model predicts scaling with the strength of feedback).  We see no statistically strong dependence of $\sigma$ on $v_c$, suggesting that gas accretion/migration is a subdominant source of gas turbulence in the FIRE disks.}
	\label{fig:vctest}
\end{figure}


There have been a number of attempts to understand the theoretical relationship between local star formation rates and the local gas disk structure (e.g., scale height, velocity dispersion).  Often, these attempts have been viewed through the lens of star formation as an equilibrium process in galaxies \citep[e.g.,][]{Ostriker2011, Faucher-Giguere2013, Hayward2017}, given the fact that the most massive stars form and live only for a fraction of a galactic dynamical time.  However, some work has sought to understand how non-equilibrium models of star formation might behave in Milky Way-like disk environments, when gas dynamical and stellar feedback timescales are on a comparable footing \citep{Benincasa2016, Orr2019}. 

Generally, models of star formation in disks either invoke feedback as a regulator of either the gravitational weight of disks or the momentum in the cold turbulent ISM or posit that star formation is solely a consequence of gas dynamics in the disk.  Usually, SFRs are tied in some way to the turbulence (velocity dispersion) in the gas disk. However, there are classes of models where star formation is the result of disk/galactic structure and acts as a regulator of $\sim$kpc-scale ISM structure and is not explicitly related to $\sigma$, for instance that of \citet{Ostriker2011}, where the feedback from stars balances the weight of the disk.  The weight of a disk being approximately described as $\frac{\pi}{2}G\Sigma_{\rm gas} (\Sigma_{\rm gas} + \gamma \Sigma_\star)$, where the $\gamma$ term is the fraction of the stellar surface density within a gas scale height.  This weight is balanced with the feedback momentum $(P/m_\star) \Sigma_{\rm SFR}$.  Thus, $\left({P}/{m_\star} \right) \Sigma_{\rm SFR} \approx \frac{\pi}{2} G \Sigma_{\rm gas} (\Sigma_{\rm gas} + \gamma \Sigma_\star)$.
This formulation for the star formation rate is agnostic to the actual velocity dispersion or scale height of the gas disk\footnote{The gas scale height ($\sim \sigma_z/\Omega$) does indirectly enter into $\gamma$, in the relative difference in scale heights of the gaseous and stellar disks.}.  Often, however, it \emph{is} connected by invoking $\tilde Q_{\rm gas} \sim \sigma \Omega/\Sigma_{\rm disk} \sim 1$, in which case the result is identical to the scaling derived from balancing feedback-injection with turbulent dissipation.  Barring that assumption, this feedback-balanced formulation actually makes \emph{no direct} prediction of the velocity dispersions or turbulence in the disk.

Among many scalings for galactic disks, there are several which are broadly argued for to explain velocity dispersions in Milky Way-like galaxies:
\begin{enumerate}

\item \label{subsec:Q1theory} \textbf{Marginal Gas Stability ($\tilde Q_{\rm gas} = 1$):} 
Gas drives itself to marginal stability against gravitational fragmentation and collapse (per the earlier discussion in \S~\ref{sec:intro}), as measured by the Toomre-Q parameter \citep{Toomre1964}.  If there is sufficient turbulent support ($\sigma$) against fragmentation, it is expected that gas would continuously dissipate turbulent energy ($\sigma$) until that gas becomes unstable and fragments, collapsing into stars.  As part of a feedback-regulation framework, it would be argued that those newly formed stars would inject momentum to drive the remaining gas back to turbulent stability, hence the fragmentation threshold of $\tilde Q_{\rm gas} = 1$ (Eq.~\ref{eq:Q}) representing a natural equilibrium solution, and predicting velocity dispersions to scale as:
\be \label{eq:sigmaQ}
\sigma_{z,(Q=1)} =  \frac{\pi G (\Sigma_{\rm gas} + \gamma\Sigma_\star)}{\sqrt{2} \Omega}  \; .
\ee


\item \label{subsec:FBtheory} \textbf{Turbulent Momentum Injection Balances Dissipation:}  
The rate of feedback momentum injected by massive stars balances the rate of turbulence dissipation in the supersonic ISM \citep[e.g.,][]{Faucher-Giguere2013, Hayward2017}.  Here, turbulent momentum is argued to decay on an eddy (disk) crossing time, where $P_{\rm turb} \sim \Sigma_{\rm gas} \sigma$ and so $\dot P_{\rm turb} \sim -\Sigma_{\rm gas} \sigma \Omega/2$.  The turbulence dissipation is balanced with feedback injection, injected at a rate of $(P/m_\star)\Sigma_{\rm SFR}$.  Together, $(P/m_\star)\Sigma_{\rm SFR} \approx \sqrt{3}\Sigma_{\rm gas}\sigma_z\Omega/2$.  Or,
\be \label{eq:sigmaFB}
\sigma_{z,FB} \approx \frac{2}{\sqrt{3}} (P/m_\star) \frac{\Sigma_{\rm SFR}}{\Omega \Sigma_{\rm gas}}  \; .
\ee
The intuition for this scaling is fairly straightforward: in order to maintain high levels of gas turbulence (velocity dispersion), large amounts of momentum (sourced from stellar feedback) need to be continually injected.  Combined with the argument that disks regulate themselves to $\tilde Q_{\rm gas} \approx 1$, this model does a fairly good job matching observations of the Kennicutt-Schmidt relation \citep{Orr2018}, and so it and the previous formulation are often conflated as the ``feedback-regulated'' model.

\item \label{subsec:CEtheory} \textbf{Constant (Low) Efficiency Star Formation:}  
Star formation proceeds at a constant efficiency per free-fall time in a gas disk \citep[e.g.,][]{Krumholz2012, Salim2015}.  Here, $\Sigma_{\rm SFR} = \epsilon_{\rm ff} \Sigma_{\rm gas} / t_{\rm ff}$, where $\epsilon_{\rm ff} \approx 0.01$, and $t_{\rm ff}$ can be derived arguing that the self-gravity of the stellar component is important, hence: $t_{\rm ff} \approx \sqrt{{3 \pi \sigma_z}/{32 G \Omega \Sigma_{\rm disk}}}$.  Combining this with previous statement of constant efficiency, we find that the star formation rate scales as $\Sigma_{\rm SFR} \propto \epsilon_{\rm ff} \Sigma_{\rm gas}\Omega^{1/2} \Sigma_{\rm disk}^{1/2} /\sigma_z^{1/2}$. And thus the velocity dispersions scale explicitly as:
\be  \label{eq:sigmaCE}
\sigma_{z,CE} \approx \frac{32 G}{3 \pi} \epsilon_{\rm ff}^2 \frac{\Omega \Sigma_{\rm gas}^2 \Sigma_{\rm disk} }{\Sigma_{\rm SFR}^2}\; .
\ee 
This scaling has the opposite intuition from the feedback-regulated model: low velocity dispersion (i.e., razor-thin) disks will have very high SFRs, and high dispersion disks will have low SFRs.  There is no explicit connection to gas stability in the form of $\tilde Q_{\rm gas} \approx 1$.

\item \label{subsec:APtheory} \textbf{Accretion Powered Turbulence/Star Formation:}
The gravitational potential energy of in-falling gas is converted to turbulent motions, i.e., $\sigma$, and gas migrates through the disk to the centers of galaxies, all while star formation is occurring \citep{Klessen2010, Krumholz2010, Cacciato2012, Krumholz2018}.  

Starting from the assumption that gas follows, albeit slowly decaying, circular orbits, the change in gravitational energy in time for a parcel of gas is: $\frac{dE}{dt} = -m v_c \frac{dv_c}{dr}\frac{dr}{dt} \approx - \beta m \frac{v_c^2}{r} v_r$, where we take $\beta$ to be an order unity parameter for the shape of the circular velocity profile in the galaxy, and $v_r$ is the radial velocity of that gas parcel.  Arguing that some fraction of this energy $\psi$ is deposited into gas turbulence, which is maintained in steady state, we require that the change in turbulent energy due to orbital decay/gas migration is thus $\psi \dot E_t \approx \psi m \sigma^3/L_{\rm eddy}$, where $L_{\rm eddy} \sim H \sim \sigma/\Omega$ is the turbulent eddy scale taken to be roughly the disk scale height.  Equating the decaying orbital energy with the turbulent energy dissipation:
\be
\beta \frac{v_c^2}{r} v_r \approx  \psi \sigma^2 \Omega \; .
\ee
And rearranging for $\sigma$, and using $\Omega = v_c/r$, yields: $\sigma^2 \approx \frac{\beta}{\psi} v_c v_r$. Arguing that the mass flux of gas through turbulent eddy-wide annuli in the disk is $\dot M_{\rm in} = 2 \pi r H \rho_{\rm gas} v_r= 2 \pi r \Sigma_{\rm gas} v_r$, we see that the velocity dispersions scale with $\dot M_{\rm in}$ as:
\be
\sigma^2 \approx {\frac{\beta}{\psi} \frac{v_c \dot M_{\rm in}}{2 \pi r \Sigma_{\rm gas}}} \; .
\ee
In steady state, some fraction of the gas entering the eddy-wide annuli is consumed in star formation, thus $\dot M_{\rm in} \approx f \dot M_{\rm SFR} = \Sigma_{\rm SFR} 2 \pi r H f = \Sigma_{\rm SFR} 2 \pi r \sigma f/\Omega$.  Substituting in for the SFR, we have:
\be \label{eq:sigmaAP}
\sigma_{z,AP} \approx \frac{\beta f}{\psi} v_c \frac{\Sigma_{\rm SFR}}{\Omega \Sigma_{\rm gas}} \; .
\ee
Barring the normalization, the accretion-powered scaling for velocity dispersion is identical to that of the feedback-regulated model (Eq.~\ref{eq:sigmaFB}) in terms of $\Sigma_{\rm SFR}/\Omega\Sigma_{\rm gas}$.

 
\end{enumerate}
\subsubsection{Comparing velocity dispersion models}
Figure~\ref{fig:FBvG} shows the result of using the local surface densities, dynamical times, etc. in the spatially resolved FIRE data to predict the velocity dispersions in the gas using these four models.  The predictions are compared against the actual velocity dispersions calculated in the simulations in both the neutral, and cold and dense gas.  For the $\tilde Q_{\rm gas} = 1$ model, we plot the predicted velocity dispersions in the pixels at their 10 Myr-averaged SFR to see essentially how well Toomre-Q alone describes the gas velocity dispersions in the pixels (it does not have a direct connection to $\Sigma_{\rm SFR}$).

In the left panel, comparing neutral velocity dispersions and gas surface densities in the FIRE simulations with these models, we see that both the feedback-regulated and $\tilde Q_{\rm gas} = 1$ models do quite well in predicting the `observed' velocity dispersions, whereas the constant-efficiency and accretion-powered models significantly over- and under-shoot the simulations.  Interestingly, the scatter seen in velocity dispersions is better reproduced by the feedback-regulated model, compared to the $\tilde Q_{\rm gas} = 1$ model, as $\Omega$ and local gas fractions do not have the same degree of variance compared with SFRs at any given gas surface density in the simulations.

The right panel of Figure~\ref{fig:FBvG} investigates the model predictions for $\sigma$ as a function of SFR, compared against the cold and dense gas in FIRE.  Here we see the stark difference in scalings between the constant-efficiency star formation model, and all others.  Given the inverse-square dependence of $\sigma$ on SFR, the constant-efficiency model greatly over-predicts (under-predicts) velocity dispersions at low (high) SFRs in significant disagreement with the simulations.  We see both how well the $\tilde Q_{\rm gas} = 1$ model predicts the dense gas velocity dispersions at all SFRs, and how the feedback-regulated and accretion-powered models scale in unison (modulo their normalizations relative to the strength of feedback $(P/m_\star)$ and circular velocities $v_c$).

In support of the feedback-regulated model, at high star formation rates, as per \S~\ref{subsec:outflows} and the work of \citet{Hayward2017}, it is likely unfair to argue that all of the momentum goes into the velocity dispersions of the cold molecular medium.  Rather, some outflowing material or heating of the ISM is warranted, and ought to carry away some of the momentum budget, and this framework naturally then overestimates the velocity dispersion in this regime.  On the other hand in this regime, the constant efficiency model significantly under-predicts $\sigma_z$, and would most easily be rectified by arguing that $\epsilon_{\rm ff}$ itself scales with $\Sigma_{\rm gas}$ (cf., \citealt{Grudic2018}).

At the other extreme, for low $\Sigma_{\rm SFR}$, the feedback regulated framework under-predicts $\sigma_z$ by $\sim$0.5 dex, with the constant efficiency model over-predicting it by about $\sim$1.5 dex.  In this case, the picture of the ISM as a purely turbulently supported medium breaks down, and the ISM begins to have significant (but not sufficient) thermal support from the atomic phase ($T \approx 8000$ K, $c_s\approx 6$ km/s).  Including the thermal component $c_s$ in quadrature with the turbulent scaling predictions would not significantly affect the prediction of the constant efficiency model, as it predicts hundreds of km/s of dispersions at the low end of star formation, but would bring the feedback-regulated model into closer agreement.

Again, as seen in the both panels of Figure~\ref{fig:FBvG}, assuming that all gas (whatever its star formation rate may be) locally has $\tilde Q_{\rm gas} = 1$ (Eq.~\ref{eq:Q}) is perhaps the best predictor of gas velocity dispersion in the cold and dense ISM phase.  And so, marginal stability against gravitational fragmentation and collapse must hold primacy over any model of resulting star formation in galaxies.

We explore the issue of degeneracy between the predictions of the feedback-regulated and accretion-powered models with Figure~\ref{fig:vctest}.  Given that both models predict $\sigma_z$ to scale with $\Sigma_{\rm SFR}/\Omega\Sigma_{\rm gas}$, but that the accretion-powered model predicts a linear scaling with $v_c$, we plot the neutral gas velocity dispersions in the FIRE simulations against their local values of $\Sigma_{\rm SFR}/\Omega\Sigma_{\rm gas}$, and color the figure by the average pixel $v_c$.  We see no statistically significant dependence of $\sigma_z$ on $v_c$, and thus conclude that though gravitational energy surely decays into \emph{some} turbulence in the simulations, that it is not the primary source of $\sigma$, being subdominant to the effects of feedback or the assumption of marginal gas stability.

In summary, both the feedback-regulated and $\tilde Q_{\rm gas} = 1$ are fairly predictive of the velocity dispersions in gas as a function of SFR in the FIRE Milky Way-like disk galaxies at kpc scales.   Accretion-powered turbulence appears to be a subdominant contributor to the overall velocity dispersions in the galaxies, and assuming that star formation proceeds at constant efficiency appear to produce strong disagreement with the velocity dispersions seen in FIRE and observations. Additionally, there are reasonable physical arguments (e.g., accounting for outflows and thermal support) that may help reduce tension at high and low star formation rates between the feedback-regulated model.  However, these physical effects only add to the difficulty of the constant-efficiency model in explaining the observed velocity dispersions.  


\section{Summary \& Conclusions}

In this paper, we explored the properties of the various spatially resolved line-of-sight gas velocity dispersions, and their relationships with different local star formation rate tracers and gas properties.  These properties were investigated in the context of face-on `observations' of the Milky Way-mass disk galaxies ($M_\star \approx 10^{10.2}-10^{10.9}$ M$_\odot$ at $z\approx 0$) in the FIRE-2 simulations, and found to be in good agreement with resolved observations of local spiral galaxies.  Our principal results are as follows:

\begin{itemize}
\item Velocity dispersions in neutral (atomic + molecular) gas are nearly constant across 3 dex in 10 Myr average star formation rates, distributed between $\approx$15-30 km/s in our sample of simulated Milky Way mass spirals (Fig.~\ref{fig:tracers}).  For both the neutral, and the cold and dense gas velocity dispersions, however, there exists a lower envelope in dispersions as a function of SFR (e.g., on 750 pc scales, we see no gas with $\sigma_z = 20$ km/s and $\Sigma_{\rm SFR} = 1$ M$_\odot$ yr$^{-1}$ kpc$^{-2}$).

\item In regions with low recent (10 Myr-avg.) star formation rates, velocity dispersions correlate with past (100 Myr-avg.) star formation rates (Fig.~\ref{fig:gasprops}).  This correlation coincides with the timescale over which past star formation events will have injected all their feedback momentum into the ISM and driven gas to relative stability ($\tilde Q_{\rm gas} \gtrsim 1$).  

\item The outer contours of the $\sigma_z$-$\Sigma_{\rm SFR}$ relation correspond to conditions where the ISM is hosting star formation rates (over different timescales) sufficiently high to expel significant fractions of the ISM as outflows/galactic fountains (Fig.~\ref{fig:outflows}).  In other words, the ISM can only sustain so much feedback over 10-100 Myr timescales, without being driven out as outflows/galactic fountains.

\item Dense gas depletion times ($\Sigma_{C\&D}/\Sigma_{\rm SFR}^{\rm 10 \, Myr}$) are shortest in the cases where either a small fraction of the ISM is dense (thus, nearly all the dense gas is involved in star formation) or where gas surface densities are high and nearly all of the ISM is cold and dense (Figs.~\ref{fig:Tdep}).

\item There is evidence for on-off cycles of star formation in the disks (Fig.~\ref{fig:Q-tdep}): less gravitationally stable patches of ISM with little past ($\sim$10-100 Myr ago) star formation have the shortest gas depletion times/most vigorous current ($\lesssim$10 Myr) star formation rates, and conversely, more gravitationally stable regions with long gas depletion times have had star formation/feedback events in the recent past (10-100 Myr ago).

\item The FIRE-2 simulations, and observations, show that regions with higher $\Sigma_{\rm SFR}$ tend to have (albeit weakly) higher $\sigma_z$, while models with constant efficiency star formation predict that such regions should have lower $\sigma_z$.  These simulations and observations are in good agreement with feedback-regulated models that predict SFRs should scale positively with velocity dispersions, and strongly disagree with predictions assuming constant local star formation efficiency.

\end{itemize}

The velocity dispersion structure of the FIRE Milky Way-mass spirals fits in with our general understanding of gas disks in the turbulently supported framework of disk structure stability.  By and large, the dispersions seen and their attendant star formation rates are consistent with the feedback-regulated model of star formation (cf.,\S~\ref{subsec:scalings}), and marginal gas stability ($\tilde Q_{\rm gas} \approx 1$).

The differing timescales traced by various proxies for star formation rates (e.g., H$\alpha$ fluxes tracing $\lesssim 10$ Myr vs. UV fluxes tracing $\lesssim 100$ Myr timescales) and the dynamical times involved for the evolution of turbulent gas line-widths (disk/eddy crossing times) are a frequent difficulty in interpreting studies of velocity dispersions and star formation rates in galaxies.  However, they also pose a unique opportunity to study the dynamical evolution of the ISM over those timescales.  We have seen evidence in the relative flatness of the $\sigma_z$-$\Sigma_{\rm SFR}$ relation for star formation rates over 10 Myr timescales that the ISM can respond relatively quickly in terms of \emph{firing}-up/turning-off star formation compared to the actual driving/decay rate of gas turbulence.  Further, in concert with measurements of the current gas stability ($\tilde Q_{\rm gas}$), simultaneously using multiple SFR tracers with different timescales can provide evidence for both feedback regulation timescales, and vigorous on-off (i.e., bursty) cycles of star formation occurring on $\sim$10-100 Myr timescales (at $\sim$kpc-scales), and differentiate star formation/galaxy disk models. 


\section*{Acknowledgements}

MEO is grateful for the encouragement of his late father, SRO, in studying astrophysics, and is supported by the National Science Foundation Graduate Research Fellowship under Grant No. 1144469.  The Flatiron Institute is supported by the Simons Foundation.  Support for AMM is provided by NASA through Hubble Fellowship grant \#HST-HF2-51377 awarded by the Space Telescope Science Institute, which is operated by the Association of Universities for Research in Astronomy, Inc., for NASA, under contract NAS5-26555.  Support for PFH was provided by an Alfred P. Sloan Research Fellowship, NASA ATP Grant NNX14AH35G, and NSF Collaborative Research Grant \#1411920 and CAREER grant \#1455342. CAFG was supported by NSF through grants AST-1412836 and AST-1517491, by NASA through grant NNX15AB22G, and by STScI through grants HST-AR-14293.001-A and HST-GO-14268.022-A. DK acknowledges support from the NSF grant AST-1412153 and Cottrell Scholar Award from the Research Corporation for Science Advancement. EQ was supported by NASA ATP grant 12-ATP12-0183, a Simons Investigator award from the Simons Foundation, and the David and Lucile Packard Foundation.




\bibliographystyle{mnras}
\bibliography{library} 



\appendix
\section{Spatially Resolved Maps and Gas Velocity Dispersion--SFR Relations of Individual Galaxy Simulations}\label{sec:appendix}

\begin{figure*}
	\centering
	\includegraphics[width=0.97\textwidth]{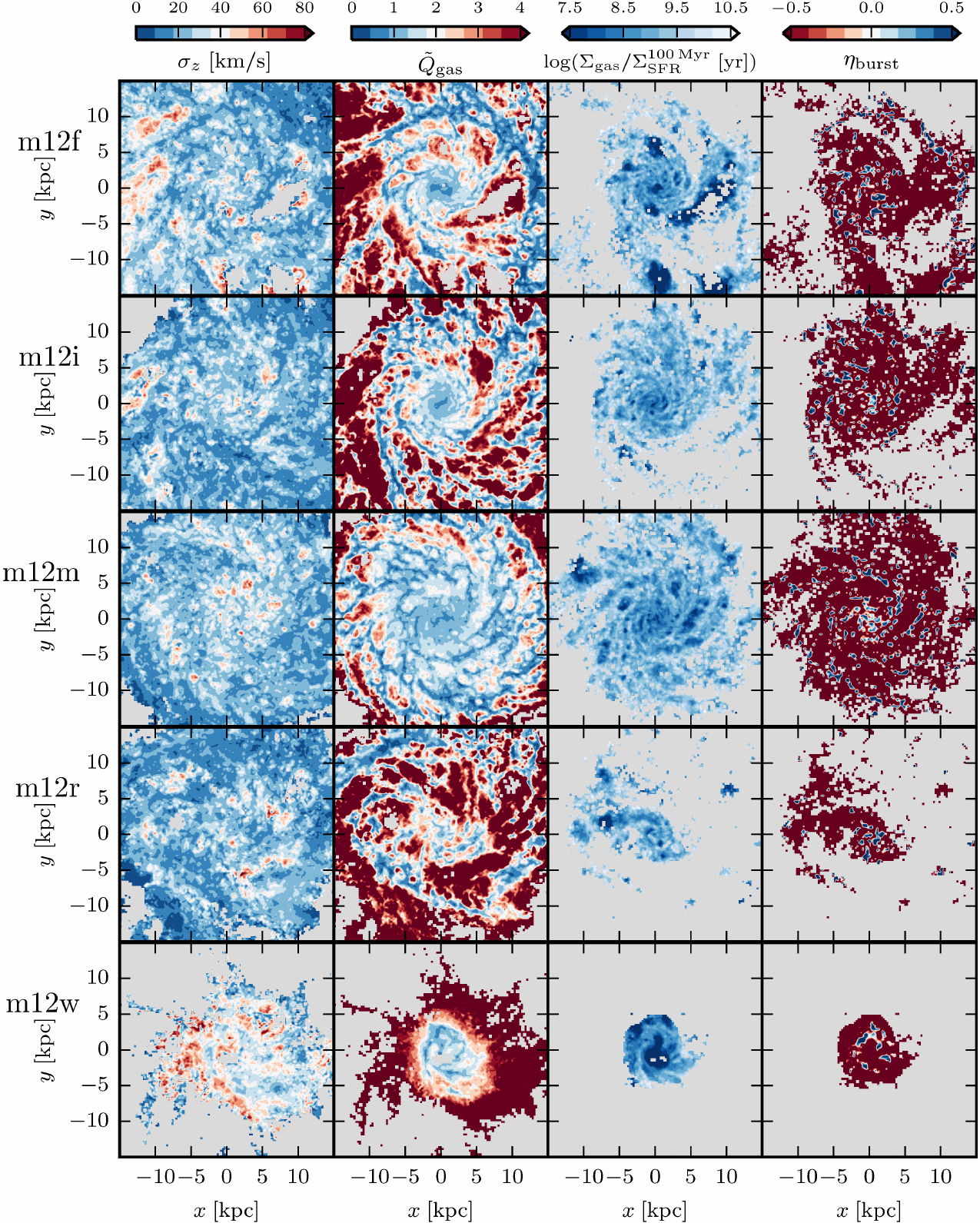}
	\caption{Identical in style and plotted quantities to Fig.~\ref{fig:galsm12b-i}, but for FIRE simulations: \textbf{m12f}, \textbf{m12i}, \textbf{m12m}, \textbf{m12r}, and \textbf{m12w} at $z= 0$. Galaxy \textbf{m12w} is the least gas rich and most compact in its gas disk of the sample.  Unlike the other more extended disks in the sample, \textbf{m12w} consists entirely (in gas) of a $\sim 3$~kpc in radius dense, gravitationally unstable, nuclear gas disk.}
	\label{fig:galsm12m-w}
\end{figure*}
\begin{figure*}
	\centering
	\includegraphics[width=0.95\textwidth]{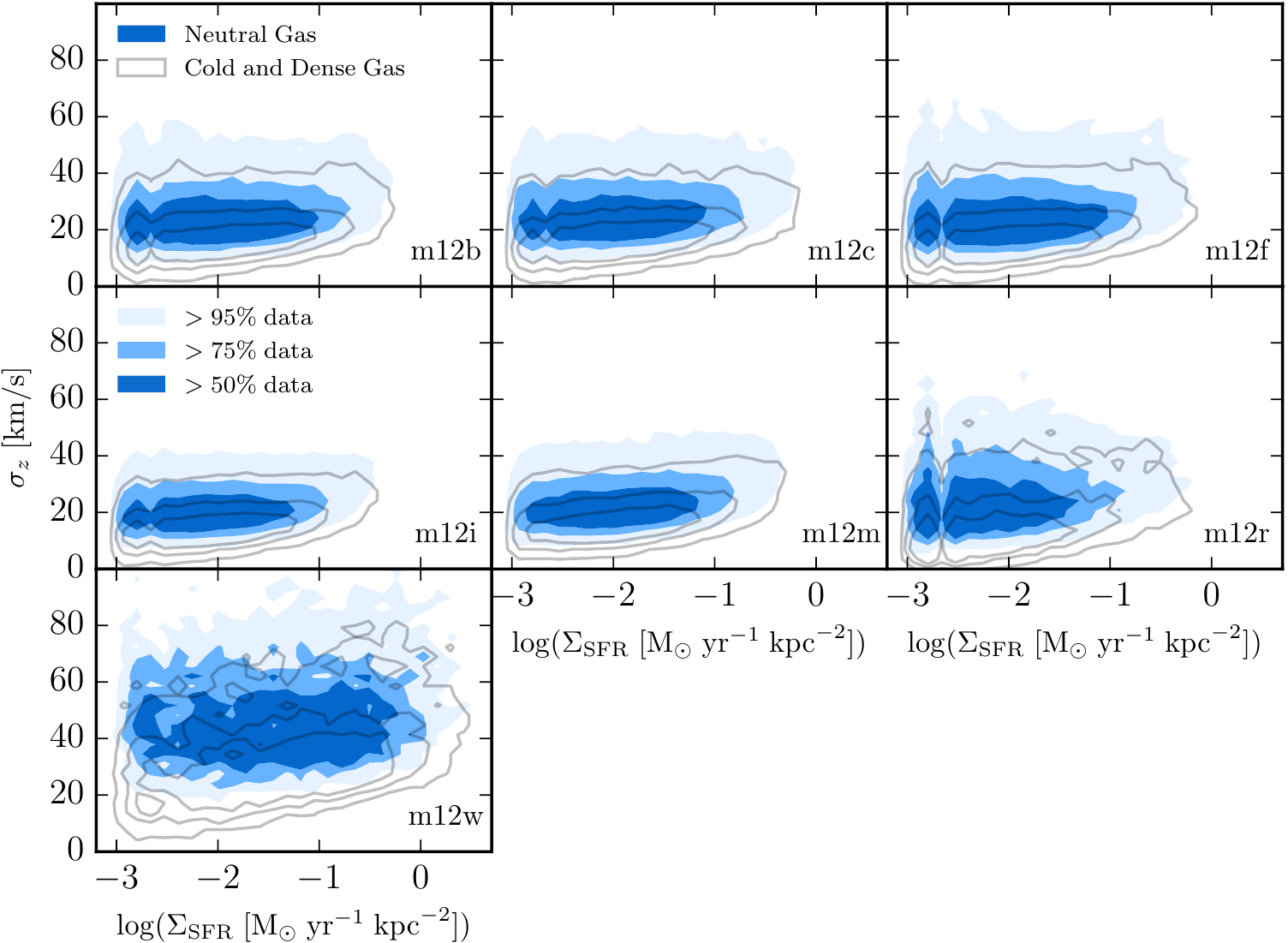}
	\caption{Distributions of spatially resolved (750 pc pixel size) line-of-sight gas velocity dispersions ($\sigma_z$) and SFR surface densities for various gas tracers in the FIRE simulations for $z \lesssim 0.1$ (as Fig.~\ref{fig:tracers}), in individual FIRE galaxies. Distributions in gas velocity dispersions and SFRs are similar across the individual galaxies, as they are all representative of face-on MW-like disk galaxies.  \textbf{m12r} and \textbf{m12w} represent the most dynamically disturbed disk systems in the core FIRE suite, and their higher-$\sigma_z$ distributions reflect this.
	}
	\label{fig:galrelations}
\end{figure*}
In total, our sample included seven Milky Way-mass disk galaxies from the FIRE-2 suite.  Figure~\ref{fig:galsm12b-i} in the main text shows spatially resolved maps of the neutral gas velocity dispersion ($\sigma_z$), Toomre-Q ($\tilde Q_{\rm gas}$), 100 Myr-averaged gas depletion time ($\Sigma_{\rm gas}/\Sigma_{\rm SFR}^{\rm 100 \, Myr}$), and star formation `burst' parameter ($\eta_{\rm burst}$) in two of these galaxies.  Figure~\ref{fig:galsm12m-w} shows the remaining five simulated galaxies.  Compared to the other disk galaxies, \textbf{m12w} is distinct in this sample, both morphologically, and having higher velocity dispersions.

Figure \ref{fig:galrelations} shows the line-of-sight velocity dispersion structure as a function of the 10 Myr-averaged star formation rate at low redshift, $z \lesssim 0.1$, on 750~pc scales in all seven individual galaxy simulations.  Two differently weighted velocity dispersions are plotted: the neutral (atomic + molecular) gas mass-weighted dispersions in (blue) colored contours, and the ``cold and dense'' ($T < 500$ K, $n_H > 1$ cm$^{-3}$) gas mass-weighted dispersions in unfilled contours.  The cold and dense gas dispersions have a very similar structure in $\sigma_z$-SFR space, albeit with a lower overall normalization indicative of its dynamically colder state.

By and large, all of the Milky Way mass spirals (those with the \textbf{m12} signifier) show similar structure in dispersion-SFR phase space.  There is little variation in the structure of the dispersions across $\sim3$~dex in SFRs, with the exception of a rising lower envelope in velocity dispersions with SFR.  \textbf{m12w} stands out with regards to the other simulations in its velocity dispersion structure.  However, visually (bottom row, Fig.~\ref{fig:galsm12m-w}), this galaxy appears morphologically distinct in its gas disk with a $\sim 3$~kpc radius gravitationally fragmenting nuclear gas disk, lacking the $\sim 10$~kpc radial extent of the other simulations at $z=0$.  To a lesser extent, \textbf{m12r} also stands out: visually, it has an irregularly structured gas disk at $z=0$ compared to the other spirals, but does not have as compact a gas distribution as \textbf{m12w}.  Given the similarity of the apparent physical structure of their disks, and their distributions in $\sigma_z$--SFR space, we believe it justified to stack together the pixels from all of the simulation snapshots into a single dataset to explore the connections between gas velocity dispersion and local SFR in MW-like disk environments in the main text.


\label{lastpage}
\end{document}